%% file: main.tex
\documentclass[twocolumn,tighten,astrosymb,trackchanges]{aastex631}

\input{affiliations}

\usepackage{CJK}
\usepackage{multirow}
\let\tablenum\relax
\usepackage{siunitx}
\usepackage{appendix}
\usepackage{amsmath,amsthm,amsfonts,amssymb,amscd}
\usepackage{booktabs}
\usepackage{tablefootnote}
\usepackage{lipsum}
\usepackage{changepage}

\let\ts=\thinspace
\newcommand{\one}{\ts {\sc i}}
\newcommand{\two}{\ts {\sc ii}}
\newcommand{\three}{\ts {\sc iii}}
\newcommand{\four}{\ts {\sc iv}}
\newcommand{\five}{\ts {\sc v}}
\newcommand{\six}{\ts {\sc vi}}
\newcommand{\seven}{\ts {\sc vii}}

\definecolor{maroon}{rgb}{0.760,0.118,0.337}

\NewDocumentCommand{\companioncitep}{o o m}{%
{\hypersetup{citecolor=red}%
\IfNoValueTF{#1}
{\citep{#3}}
{\IfNoValueTF{#2}
{\citep[#1]{#3}}
{\citep[#1][#2]{#3}}}}%
}

\NewDocumentCommand{\companioncitet}{o o m}{%
{\hypersetup{citecolor=red}%
\IfNoValueTF{#1}
{\citet{#3}}
{\IfNoValueTF{#2}
{\citet[#1]{#3}}
{\citet[#1][#2]{#3}}}}%
}

\def\cm{\mbox{\,cm}}
\def\cm3{\mbox{\,cm$^{-3}$}}

\shorttitle{Lens modelling of SN 2025wny}
\shortauthors{Mörtsell et al.}
\begin{document}

\title{
Follow-up of SN 2025wny V: Lens Modelling and Cosmography of a Strongly Lensed Superluminous Supernova at $z = 2.015$ using Space Data
}

\correspondingauthor{Edvard~Mörtsell}
\email{edvard@fysik.su.se}
\input{authors_25wny}

%=======================================================================
\begin{abstract}

We present a lensing and cosmographic analysis of the strongly lensed Type~I superluminous supernova SN~2025wny at redshift $z=2.015$, multiply imaged by two foreground galaxies at $z=0.376$. Using imaging obtained with the {\it Hubble Space Telescope} and the {\it James Webb Space Telescope}, we model the lens system with two elliptical power-law mass distributions and an external shear component. In addition to the supernova image positions, the modelling incorporates the surface-brightness distribution of the lensed host galaxy.

The inferred Einstein radii are $\theta_{\rm E,1}\simeq1.6\arcsec$ and $\theta_{\rm E,2}\simeq0.7$--$0.8\arcsec$, with broadly consistent results across all filters. After accounting for microlensing by stars in the lens galaxies, the posterior distribution spans total magnifications of approximately $\mu_{\rm tot}\sim 5$--$50$, with flux ratios of the multiple images consistent with observations.

Combining the lens models with spectroscopically and photomerically measured time delays yields a filter-marginalized constraint of
\[
H_0 = 66.7^{+7.6}_{-6.3}\;
{\rm km\,s^{-1}\,Mpc^{-1}},
\]
for a fiducial model with isothermal mass profiles. Allowing the density slopes of the lens galaxies to vary over a broad range results in
\[
H_0 = 70.8^{+8.2}_{-6.1}\;
{\rm km\,s^{-1}\,Mpc^{-1}}.
\]

These values are conditional on the adopted parameterization of the lens
mass distribution, the assumed priors on the density slopes, and possible
additional lensing contributions from the surrounding large-scale
environment. We find that incorporating the currently available stellar
kinematic measurements has only a modest effect on the inferred value of
$H_0$. Future measurements of the lens-galaxy kinematics and a detailed
characterization of the lens environment will further strengthen the
utility of SN~2025wny as a cosmological probe.

\end{abstract}
%=======================================================================

\vspace{2.0cm}
%=======================================================================
\section{Introduction \label{sec:intro}}
%=======================================================================

Gravitationally lensed supernovae provide unique probes of both astrophysics and cosmology. In particular, they enable direct studies of galaxy-scale mass distributions and offer independent measurements of the Hubble constant, $H_0$, through time-delay cosmography \citep{refsdal_1964b}. Over the past decade, the sample of known lensed supernovae (SNe) has grown from isolated prototypes to a small but increasingly diverse population spanning a range of SN types and redshifts.

The first identified gravitationally lensed SN, PS1-10afx, was discovered serendipitously and initially classified as an unusually luminous SN \citep{2013ApJ...767..162C}. It was later shown by \citet{quimby_extraordinary_2013} to be a highly magnified normal Type Ia SN (SNIa). However, the event had faded before multiple images could be spatially resolved. The first system in which multiple images of a single SN were observed, SN Refsdal, demonstrated the characteristic appearance of a lensed transient, with several images produced by a galaxy cluster lens \citep{kelly_2015, Kelly_2016}. The first resolved, multiply imaged SNIa in a galaxy-scale lens, iPTF16geu, further highlighted the ability of wide-field time-domain surveys to discover such rare systems \citep{goobar_2017}.

More recent discoveries have expanded the sample, including the strongly magnified SNIa Zwicky \citep{goobar_2023} and additional events identified through systematic searches with the {\it Hubble Space Telescope} ({\it HST}) and the {\it James Webb Space Telescope} ({\it JWST}) \citep{rodney_2021, pierel_2024, frye_2024}. A recent addition to the sample is the strongly lensed Type II SN~2025mkn at redshift $z=1.371$, lensed by a foreground galaxy at redshift $z=0.42$ \citep{Lemon2026_SN2025mkn}.

SN~2025wny, at $z=2.015$, is a Type I superluminous supernova (SLSN-I) that is strongly lensed and multiply imaged by two foreground galaxies at $z=0.376$ \citep{Johansson:2025zpy, 2026A&A...710A.365T}. It is the first known lensed SLSN and the first SN of any type for which multiple images can be individually resolved in seeing-limited ground-based imaging. This system significantly extends the redshift reach of ground-based lensing studies and adds a new class of objects to the sample of gravitationally lensed transients. SN image positions and time delays, together with the lensed host arc, provide strong constraints on the lens mass distribution and enable an independent determination of $H_0$ through time-delay cosmography.

%=======================================================================
\section{Data \label{sec:data}}
%=======================================================================

Here we provide a brief summary of the observations leading to the identification of SN~2025wny as a strongly lensed SN. A more complete description of the discovery, follow-up observations, and space-based imaging data is presented by \companioncitet{Goobar2026wny}. The physical properties of SN~2025wny are discussed in detail by \companioncitet{Li2026wny}.

%-----------------------------------------------------------------------
\subsection{Discovery and spectroscopy}
%-----------------------------------------------------------------------

SN~2025wny was first identified by the Zwicky Transient Facility (ZTF; \citealt{Bellm+2019,Graham+2019,Dekany+2020,Masci+2019,Patterson+2019,Mahabal+2019,Duev+2019}) on UT 2025 August 29 and assigned the identifier ZTF25abnjznp.

The transient is located in close projection to a luminous galaxy with an archival spectrum from the Dark Energy Spectroscopic Instrument (DESI), which places the galaxy at a redshift of $z=0.3754 \pm 0.0001$ \citep{desi2025}. The observed brightness of the transient is inconsistent with that expected for a SN at this redshift, suggesting a more distant source subject to substantial gravitational magnification. 

Additional evidence for lensing came from the proximity of the transient to the known lens candidate PS1J0716+3821 \citep{2020A&A...644A.163C}, identified through cross-matching with the Strong Lensing Database (SLED). Archival Legacy Survey imaging \citep{2019AJ....157..168D} reveals a second nearby galaxy, while Canada-France-Hawaii Telescope (CFHT) imaging \citep{Gwyn2008PASP..120..212G} shows four images of a background source arranged in a cross-like configuration around the lens galaxies together with extended arc-like emission.

Spectroscopic observations obtained with Keck/LRIS are broadly consistent with a Type I superluminous supernova (SLSN-I), leading to the classification of SN~2025wny as a strongly lensed SLSN-I \citep{Johansson:2025zpy, 2026A&A...710A.365T}.
Integral-field spectroscopy obtained with the JWST Near Infrared
Spectrograph yields a host-galaxy redshift of $z_{\rm host}=2.0151\pm0.0001$, 
establishing the redshift of the source \companioncitep{Goobar2026wny}.

%-----------------------------------------------------------------------
\subsection{{\it HST} and {\it JWST} imaging \label{sec:imaging}}
%-----------------------------------------------------------------------

The system was subsequently observed with both {\it HST} and {\it JWST}. The {\it HST} observations comprise imaging in the optical filters F475W, F625W, and F814W, together with near-infrared imaging in F160W. The {\it JWST} observations were obtained in the F115W, F150W, and F277W filters.

\begin{figure*}[htbp]
\centering
\includegraphics[width=0.49\linewidth]{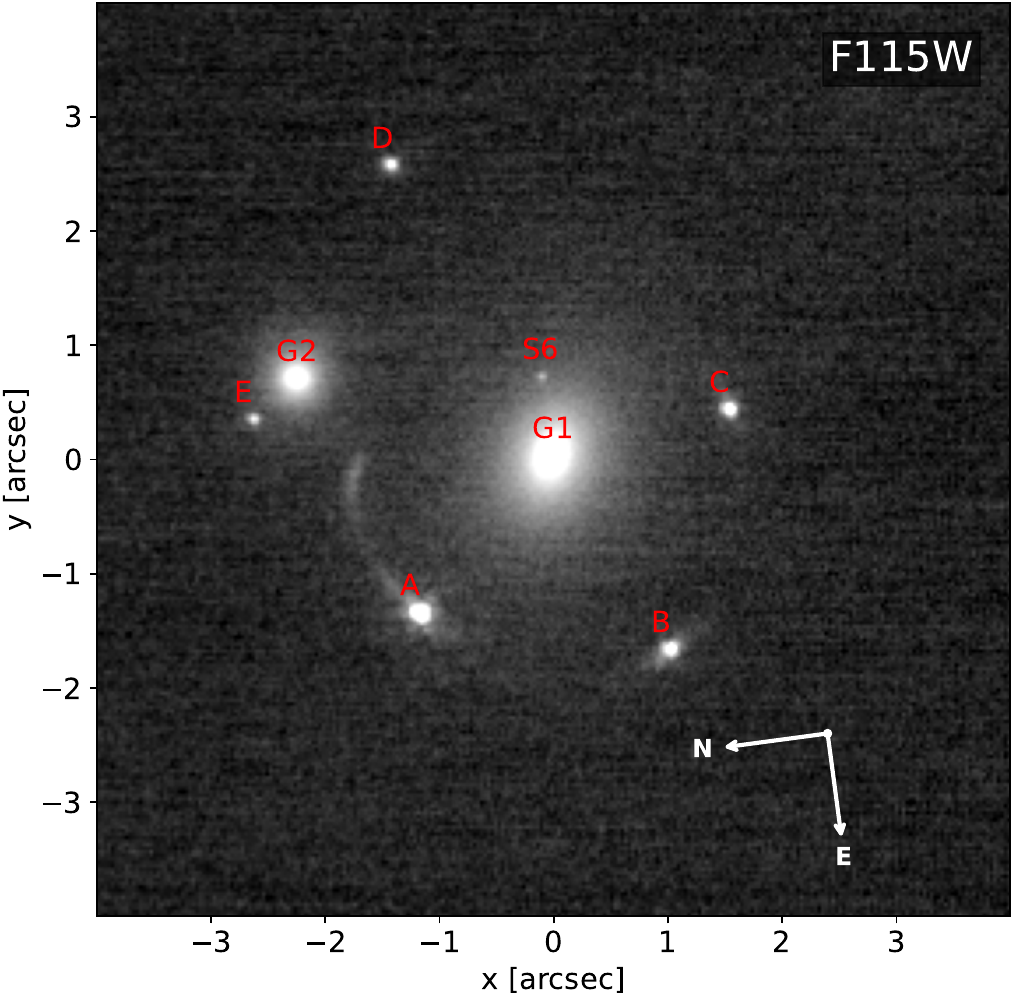}
\hfill
\includegraphics[width=0.49\linewidth]{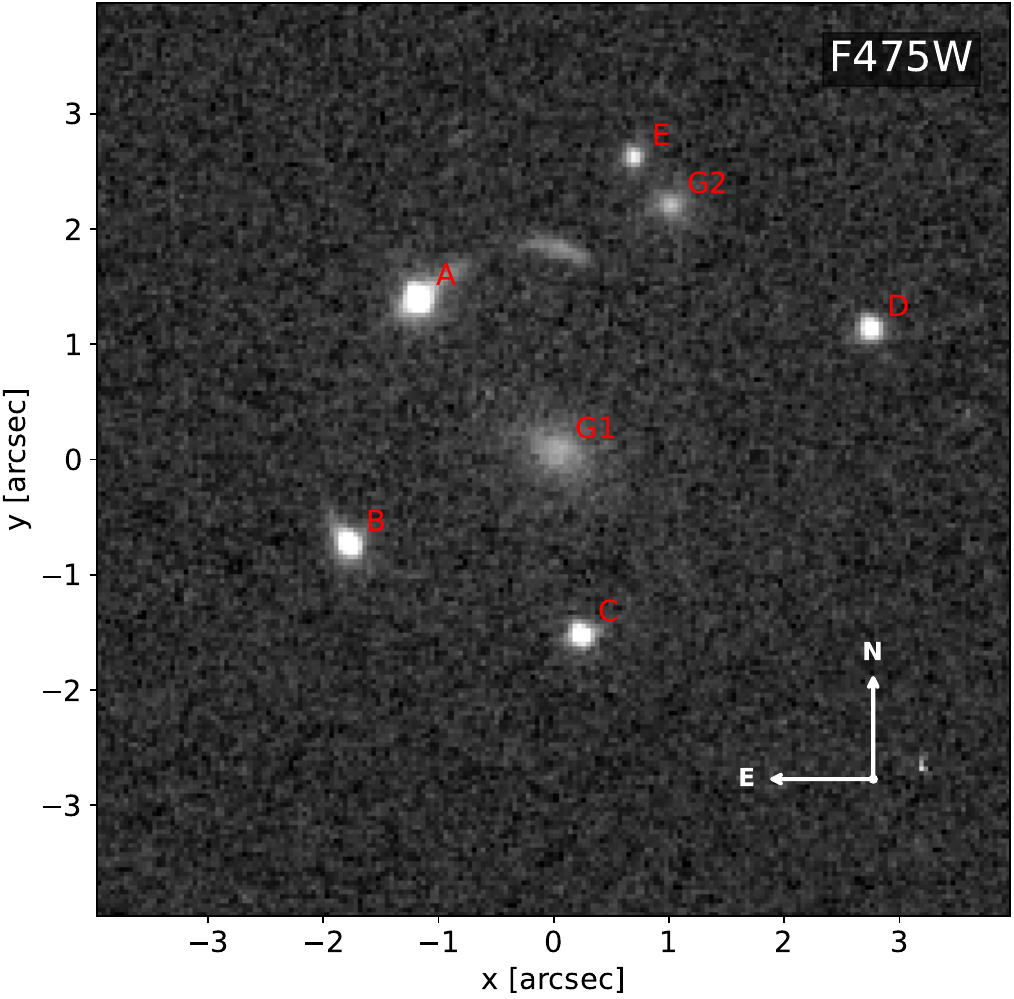}
\caption{
Imaging of the SN~2025wny system obtained with {\it JWST} F115W (left) and {\it HST} F475W (right). The five lensed images of SN~2025wny, labeled A--E, are visible around the two foreground lens galaxies G1 and G2. Extended arc-like emission from the host galaxy is detected in both datasets and is particularly prominent in the {\it JWST} imaging. An additional source, S6, is detected in the {\it JWST} image but is unlikely to be associated with the SN~2025wny lens system \companioncitep{Goobar2026wny}. The {\it JWST} image is shown in the native detector orientation to preserve the original pixel sampling.
\label{fig:imaging}
}
\end{figure*}

Figure~\ref{fig:imaging} shows the {\it JWST} F115W and {\it HST} F475W
images of the system. Five lensed images of SN~2025wny, labelled A--E,
are clearly visible around the two foreground lens galaxies, denoted G1
and G2. Extended arc-like emission from the lensed host galaxy is also
detected. In the remainder of the paper, we adopt a common lens redshift
of $z=0.376$, since the small difference between
$z_{\rm G1}=0.3755\pm0.0001$ and $z_{\rm G2}=0.3766\pm0.0001$
\companioncitep{Johansson2026wny} can plausibly be attributed to their relative
peculiar velocities.

An additional source, labeled S6 in Figure~\ref{fig:imaging}, is detected in the {\it JWST} imaging and in the {\it HST} F160W data. Its position and properties are inconsistent with the lensing configuration inferred for SN~2025wny, and it is therefore unlikely to be associated with the system \companioncitep{Goobar2026wny}.

For the lens modelling presented below, we primarily use the {\it JWST} F115W and F150W observations, which provide a pixel scale of $0.031\arcsec$\,pixel$^{-1}$. These data offer substantially higher spatial resolution than the F277W imaging ($0.063\arcsec$\,pixel$^{-1}$) and the {\it HST} F160W observations ($0.13\arcsec$\,pixel$^{-1}$). The optical {\it HST} filters F475W, F625W, and F814W ($0.040\arcsec$\,pixel$^{-1}$) are used primarily as an independent consistency check and to assess systematic uncertainties in the lens modelling.

%=======================================================================
\section{Modelling \label{sec:lensmodel}}
%=======================================================================

%-----------------------------------------------------------------------
\subsection{Light model}\label{sec:lightmodel}
%-----------------------------------------------------------------------

The light distribution of the primary lens galaxy (G1) is modelled using a combination of a S\'ersic profile and an additional exponential component (equivalent to a S\'ersic profile with fixed index $n=1$). This two-component model provides sufficient flexibility to reproduce both the central light distribution and the extended stellar envelope of the galaxy. The secondary lens galaxy (G2) is modelled with a single S\'ersic profile. We also include a spatially uniform background component to account for residual sky emission.

The host galaxy of SN~2025wny is modelled in the source plane using a single S\'ersic profile. The resulting source-plane light distribution is ray-traced through the lens model and compared to the observed host-galaxy arcs in the image plane.

The lensed SN images are treated as point sources convolved with the point-spread function (PSF) appropriate for each instrument and filter. For the {\it HST} data, the initial PSF model is constructed from the relatively isolated image D, which is expected to suffer the least contamination from host-galaxy light. For the {\it JWST} data, the initial PSF are drawn from the JWST PSF simulation library\footnote{\url{https://stsci.app.box.com/v/jwst-simulated-psf-library}}. In all filters, the PSFs are re-centered and iteratively refined during the fitting procedure using the PSF reconstruction algorithm implemented in \texttt{lenstronomy} \citep{Birrer2018,Birrer2021}.

%-----------------------------------------------------------------------
\subsection{Lens mass model}
%-----------------------------------------------------------------------

We model the mass distributions of the two lens galaxies using elliptical power-law mass profiles, for which the three-dimensional density distribution scales as
\[
\rho \propto r^{-\eta}.
\]
In the most general case, each lens is parameterized by its Einstein radius $\theta_{\rm E}$, logarithmic density slope $\eta$, ellipticity components $(e_{1,\rm lens}, e_{2,\rm lens})$, and centroid position $(x_{\rm lens}, y_{\rm lens})$. The ellipticity components are related to the axis ratio $q$ and the position angle $\phi$ through
\[
e_{\rm lens} = \frac{1-q}{1+q},
\qquad
\phi = \tfrac{1}{2}\arctan2(e_{2,\rm lens}, e_{1,\rm lens}),
\]
with
\[
e_{\rm lens} = \sqrt{e_{1,\rm lens}^2 + e_{2,\rm lens}^2}.
\]
All angles are defined in the unrotated telescope image frame.

The environmental contributions to the lensing potential are modelled through an external shear component parameterized by its amplitude $\gamma_{\rm ext}$ and orientation $\psi_{\rm ext}$.

Broad, weakly informative priors are adopted for all model parameters.
The only exception is the lens mass centroids, which are constrained to
lie within $0.1\arcsec$ ($\sim500$\,pc) of the corresponding light
centroids. Although galaxy-scale strong lenses generally exhibit much
smaller mass--light offsets \citep{Bolton2008}, we adopt this broader prior to allow for possible perturbations arising
from the close lens pair.

The primary constraints on the lens model are provided by the positions of the five SN images, yielding ten astrometric constraints. Our fiducial model adopts fixed isothermal density slopes, $\eta_1=\eta_2=2$, consistent with measurements of early-type strong-lens galaxies \citep{Shajib2021}. The resulting model contains 12 free parameters, four of which correspond to the lens mass centroids and are subject to the positional priors described above.

Modelling of the {\it JWST} imaging, where the lensed host-galaxy emission is most clearly detected, indicates that the SN is offset from the centre of its host galaxy by only $0.016^{+0.012}_{-0.007}\arcsec$ in the source plane. Consequently, the contribution of the lensed host galaxy to the observed flux at the SN image positions is non-negligible and is included in all lens-model fits. For {\it HST} imaging, where the host-galaxy arcs are significantly fainter, the relative position of the SN and host centre cannot be reliably constrained from the data alone. We therefore fix the SN source position to coincide with the centre of the host-galaxy light distribution. Although the measured source-plane offset is small, this approximation may contribute to the filter-to-filter scatter in the inferred lens parameters and the corresponding values of $H_0$.

The extent to which the density slopes can be constrained depends on both the information content of the lensed host-galaxy images and the adopted source-light model. To assess the impact of the mass-profile slope on the inferred value of $H_0$, we also perform fits in which the density slopes $\eta_1$ and $\eta_2$ are allowed to vary. These fits are restricted to the {\it JWST} F115W and F150W observations, where the host-galaxy emission is best resolved. The full set of lens mass model parameters is summarized in Table~\ref{tab:lens_params}.

For each posterior sample, we compute a range of derived lensing quantities. The image magnifications, arrival times, convergence $\kappa$, and shear amplitude $|\gamma|$ are evaluated directly at the fitted image positions associated with that realization of the lens model. Separately, we solve the lens equation for the corresponding source position to determine whether the model predicts additional images beyond those included in the fit. In this way, the full posterior distribution of the lens-model parameters, together with all parameter covariances, is propagated into the derived lensing observables.

\begin{table}[htbp]
\centering
\caption{Lens mass model parameter. Our fiducial model adopts fixed isothermal density slopes, $\eta_i=2$.}
\label{tab:lens_params}
\begin{tabular}{ll}
\toprule
Parameter & Description \\
\midrule
$\theta_{\rm E,i}$ & Einstein radius of lens galaxy $i$ \\
$\eta_i$ & Density slope, $\rho \propto r^{-\eta_i}$ \\
$e_{1,i}, e_{2,i}$ & Ellipticity components of lens galaxy $i$ \\
$x_i, y_i$ & Lens mass centroid \\
$\gamma_{\rm ext}$ & External shear amplitude \\
$\psi_{\rm ext}$ & External shear position angle \\
\bottomrule
\end{tabular}
\end{table}

%-----------------------------------------------------------------------
\subsection{Microlensing}
%-----------------------------------------------------------------------

In addition to the macroscopic magnification produced by the smooth lens potential, stars in the lens galaxies introduce microlensing, which can further magnify or demagnify the individual SN images. Microlensing primarily affects the observed fluxes and flux ratios, while producing negligible image position and  time-delay perturbations.

Previous studies of lensed SNe have shown that microlensing can produce significant deviations from the magnifications predicted by smooth lens models and must therefore be accounted for when interpreting image brightnesses (e.g., \citealt{Arendse2025}). In this work, we incorporate microlensing by convolving the macromodel magnification posteriors with microlensing magnification probability distribution functions (PDFs) derived from dedicated simulations. The details of this procedure, including the estimation of the stellar mass fraction at the image positions and the microlensing simulations, are described in Appendix~\ref{appsec:microlensingmodelling}.

%=======================================================================
\section{Results}
%=======================================================================

%-----------------------------------------------------------------------
\subsection{Previous Lens Modelling Results}
%-----------------------------------------------------------------------

The first lens models of SN~2025wny were presented by
\citet{2026arXiv260216620E}, based on adaptive-optics imaging obtained
with the Large Binocular Telescope. The lens mass distribution was
described by a singular isothermal ellipsoid (SIE) for the primary lens
galaxy G1, a singular isothermal sphere (SIS) for the secondary galaxy
G2, and an external shear component. Constrained primarily by the
astrometric positions of the five observed supernova images, the
preferred model yielded Einstein radii of
$\theta_{\rm E,1}\simeq1.6\arcsec$ and
$\theta_{\rm E,2}\simeq0.75\arcsec$, together with substantial
magnifications of the multiple supernova images.

An independent analysis was subsequently presented by
\citet{2026arXiv260402418S}, using adaptive-optics imaging obtained with
NIRC2 on the Keck telescope. Adopting a broadly similar lens model and
modelling methodology, they likewise found an isothermal mass
distribution to provide a satisfactory description of the system and
obtained lens parameters in good agreement with those of
\citet{2026arXiv260216620E}. Their analysis also explored the
constraints imposed by the absence of additional supernova images beyond
the five observed images.

Both previous studies were necessarily limited by the available
ground-based imaging and relied primarily on the astrometric positions
of the supernova images as lensing constraints. In the present work, we
revisit the lens modelling of SN~2025wny using the deep, high resolution imaging obtained with {\it HST} and {\it JWST}. In addition to the
supernova image astrometry, we jointly model the extended
surface-brightness distribution of the lensed host galaxy, providing
significantly stronger constraints on the lens mass distribution. We
also investigate the impact of allowing the density slopes of the lens
galaxies to vary and derive corresponding constraints on the Hubble
constant from the measured time delays.

%-----------------------------------------------------------------------
\subsection{New {\it HST/JWST} Lens Model}
%-----------------------------------------------------------------------

As described in Section~\ref{sec:lensmodel}, we fit lens models independently to the {\it JWST} F115W and F150W imaging, as well as the {\it HST} F475W, F625W, and F814W  observations. Our fiducial analysis adopts isothermal density profiles for both lens galaxies, corresponding to $\eta_1=\eta_2=2$. 

In addition, for the {\it JWST} F115W and F150W data we explore models in which the density slopes are allowed to vary over the interval $\eta_{1,2}\in[1.6,2.4]$ in order to assess the impact of the mass-profile slope on the inferred value of $H_0$. This interval was chosen to encompass a broad range of physically plausible density profiles around the isothermal case, while remaining substantially wider than the priors typically adopted in strong-lensing analyses. Unless otherwise stated, the results presented in this section refer to the fiducial isothermal model.

For each filter, we first model and subtract the lens-galaxy light using the procedure described in Section~\ref{sec:lightmodel}. The lens mass model is then constrained using both the SN images and the lensed host-galaxy surface brightness. We verified that simultaneously fitting the SN and lens light and lens mass model yields consistent results, albeit at substantially higher computational cost.

\begin{figure}[htbp]
\centering
\includegraphics[width=\linewidth]{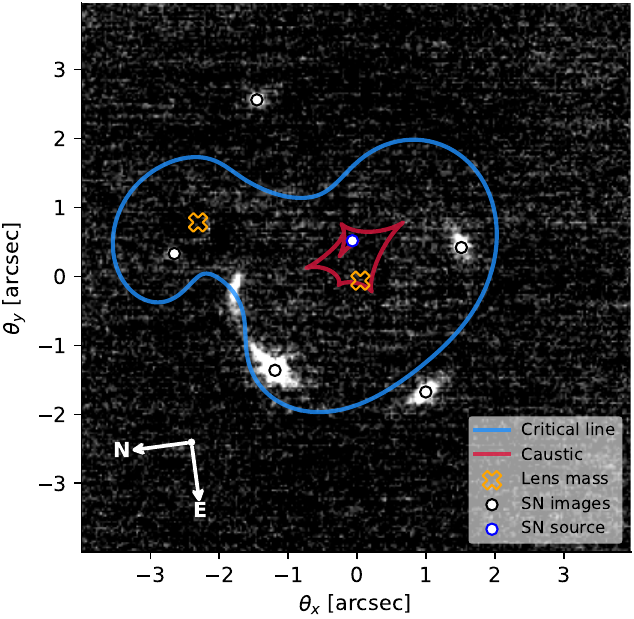}
\caption{
Best-fitting lens model for the {\it JWST} F115W data assuming isothermal density profiles ($\eta_1=\eta_2=2$). The blue curve shows the critical line in the image plane, while the red curve shows the corresponding caustic in the source plane. Orange crosses indicate the inferred mass centroids of the two lens galaxies, and open circles mark the observed positions of the five SN images. The blue point denotes the reconstructed source-plane position of SN~2025wny. The lens-galaxy light has been subtracted from the background image for clarity.
\label{fig:F115W_gamma=2_lens_model}}
\end{figure}

Figure~\ref{fig:F115W_gamma=2_lens_model} illustrates the best-fitting F115W lens model, showing the critical curves and caustics together with the inferred lens-galaxy mass centroids, the observed SN image positions, and the reconstructed source-plane position of SN~2025wny.

Lens modelling in the F115W {\it JWST} filter, constrains the lens mass within the Einstein radii of G1 and G2 to $M_1(<\theta_{\rm E,1})=(4.62^{+0.08}_{-0.07})\cdot 10^{11}\,M_\odot$ and $M_2(<\theta_{\rm E,2})=(1.05^{+0.02}_{-0.12})\cdot 10^{11}\,M_\odot$ ($95\,\%$ confidence level), respectively, with very similar numbers for F150W data.

The inferred lens-model parameters, summarized in Table~\ref{tab:lens_mass}, are highly consistent across all filters. The primary lens galaxy is found to have an Einstein radius of $\theta_{\rm E,1}\simeq1.6\arcsec$, while the secondary lens has $\theta_{\rm E,2}\simeq0.7$--$0.8\arcsec$. These values are in excellent agreement with the earlier image-position-based modelling of \citet{2026arXiv260216620E} and \citet{2026arXiv260402418S}. The lens galaxies are inferred to be moderately elliptical, with axis ratios $q_1 \simeq 0.7\text{--}0.8$ and $q_2\approx0.8$. All filters require a substantial external shear contribution, with $\gamma_{\rm ext} \simeq 0.1\text{--}0.15$, indicating a non-negligible tidal contribution to the overall lensing
potential.

The corresponding macromodel magnification predictions presented in Table~\ref{tab:magnification}, are also broadly consistent across all filters. The brightest image, A, is predicted to have a magnification of approximately $|\mu_{\rm A}|\simeq 8\text{--}14$, while the faintest image, E, has $|\mu_{\rm E}|\simeq 1.5\text{--}2$. The total magnification of the SN is substantial, with $\mu_{\rm tot}\simeq 24\text{--}31$ depending on the filter. These values are broadly consistent with the earlier lens modelling. The F814W data yield systematically larger magnifications than the remaining filters, although the inferred image ordering and relative magnifications remain unchanged.

Including microlensing substantially modifies the inferred magnification distributions, as illustrated by the microlensing probability distributions shown in Figure~\ref{fig:Microlens_PDFs}. In addition to broadening the uncertainties, the strongly skewed microlensing PDFs shift the posterior medians toward lower magnifications, as evident from Table~\ref{tab:magnification_microlensing}. The effect is most pronounced for image~A, for which the magnification posterior develops a pronounced tail toward lower magnifications.

\begin{figure}[htbp]
\centering
\includegraphics[width=\linewidth]{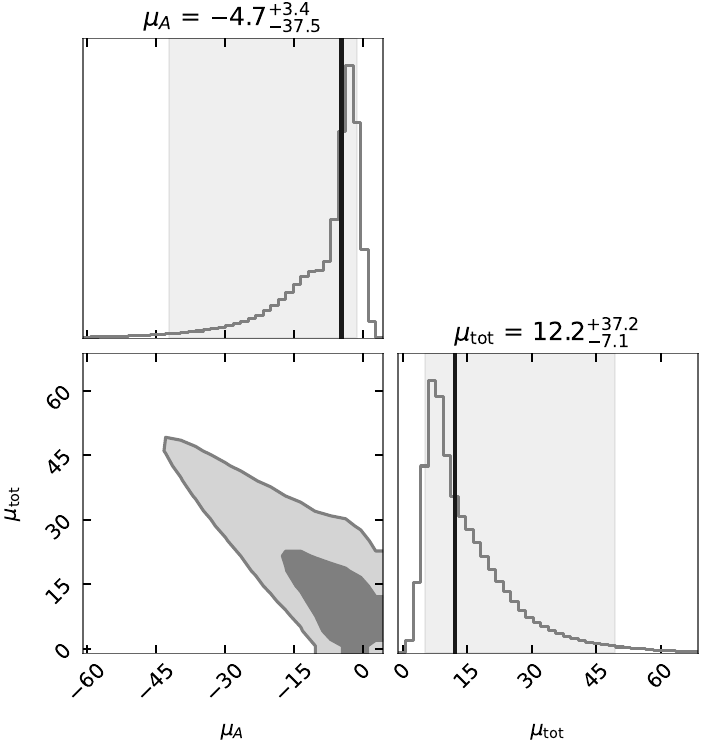}
\caption{
Posterior distributions of the magnification of image A, $\mu_{\rm A}$, and the total magnification of SN~2025wny, $\mu_{\rm tot}$, obtained from the {\it JWST} F115W lens model after incorporating microlensing. The one-dimensional panels show the marginalized posterior distributions, with vertical lines indicating the median values and shaded regions denoting the central $95\,\%$ credible intervals. The lower-left panel shows the corresponding joint posterior distribution. Microlensing introduces strongly non-Gaussian and asymmetric uncertainties, particularly for the highly magnified image A, leading to extended tails toward higher absolute magnifications.
\label{fig:A_and_tot_microlensed}}
\end{figure}

The total magnification of the system is correspondingly reduced, with median values of $\mu_{\rm tot}\simeq11$--$17$, depending on the filter, compared to $\mu_{\rm tot}\simeq24$--$31$ for the macrolens-only model. Figure~\ref{fig:A_and_tot_microlensed} shows the resulting posterior distributions for the magnification of image~A and for the total system magnification. In both cases, the uncertainties are dominated by microlensing rather than by the macrolens model, giving rise to markedly non-Gaussian posterior distributions. 

\begin{figure*}[htbp]
\centering
\includegraphics[width=0.8\linewidth]{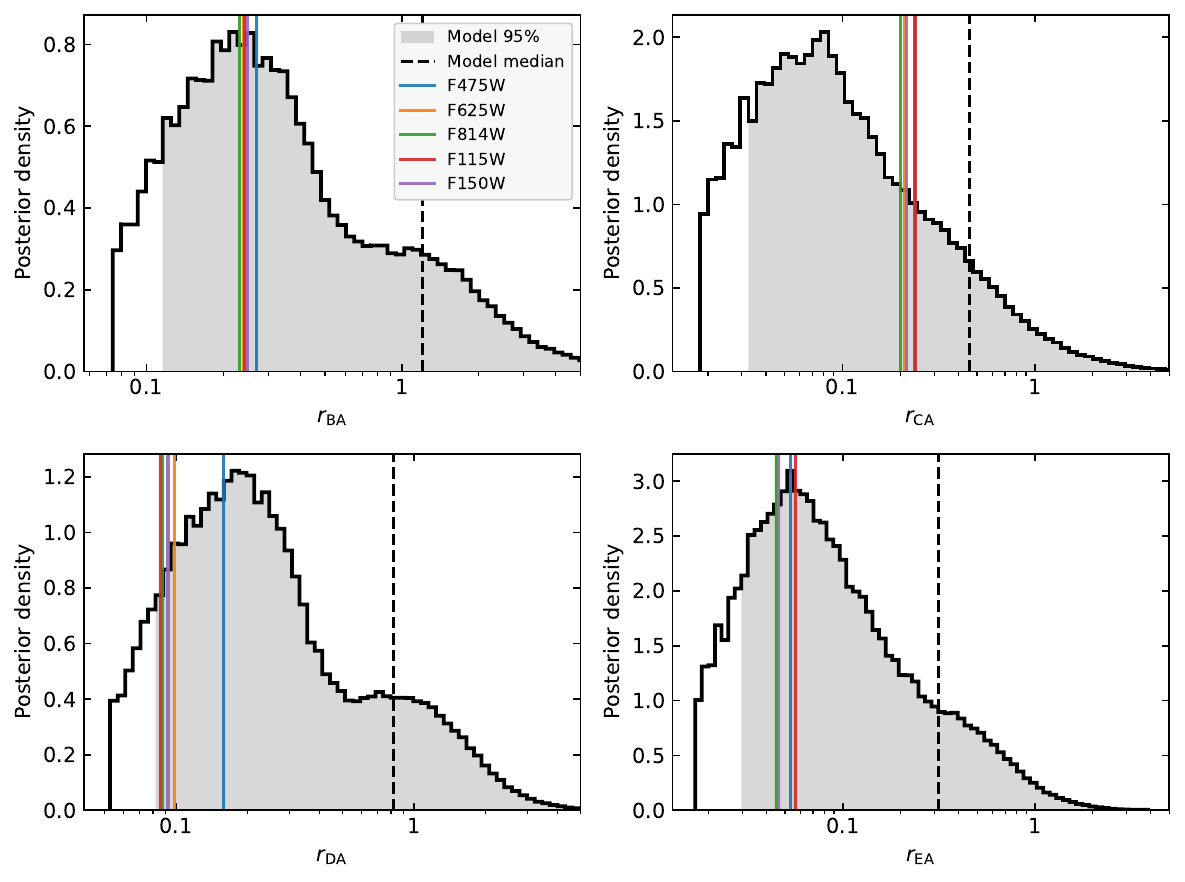}
\caption{
Comparison of observed and modelled flux ratios relative to image A. The black histogram shows the posterior distribution predicted by the combined macro- and microlensing model, while the shaded region denotes the central $95\,\%$ credible interval and the dashed vertical line marks the posterior median. Colored vertical lines indicate the observed flux ratios measured in the five filters, with the corresponding (very thin) shaded bands showing the $1\sigma$ measurement uncertainties. The horizontal axis is displayed on a logarithmic scale. All observed flux ratios are consistent with the model predictions at the $95\,\%$ confidence level, although the observed ratios are generally lower than the posterior medians, particularly for image D.
\label{fig:Ratios}}
\end{figure*}

Table~\ref{tab:ratios} and Figure~\ref{fig:Ratios} compare the observed flux ratios of images B--E relative to image A with the predictions of the combined macro- and microlensing model. Here, $f_{\rm X}/f_{\rm A}$ denotes the observed flux ratio, while $\mu_{\rm X}/\mu_{\rm A}$ denotes the model prediction. The model distributions are highly asymmetric and are therefore shown on a logarithmic scale. All observed flux ratios are consistent with the model predictions at the $95\,\%$ confidence level. However, the observed ratios are systematically lower than the posterior medians, with the largest discrepancy occurring between $f_{\rm D}/f_{\rm A}$ and $\mu_{\rm D}/\mu_{\rm A}$. This behaviour is expected, since the substantial time delay between images A and D implies that the observed ratio compares the SN at different phases of its light-curve evolution, whereas the lensing model predicts a static flux ratio.

\begin{figure*}[htbp]
\centering
\includegraphics[width=\linewidth]{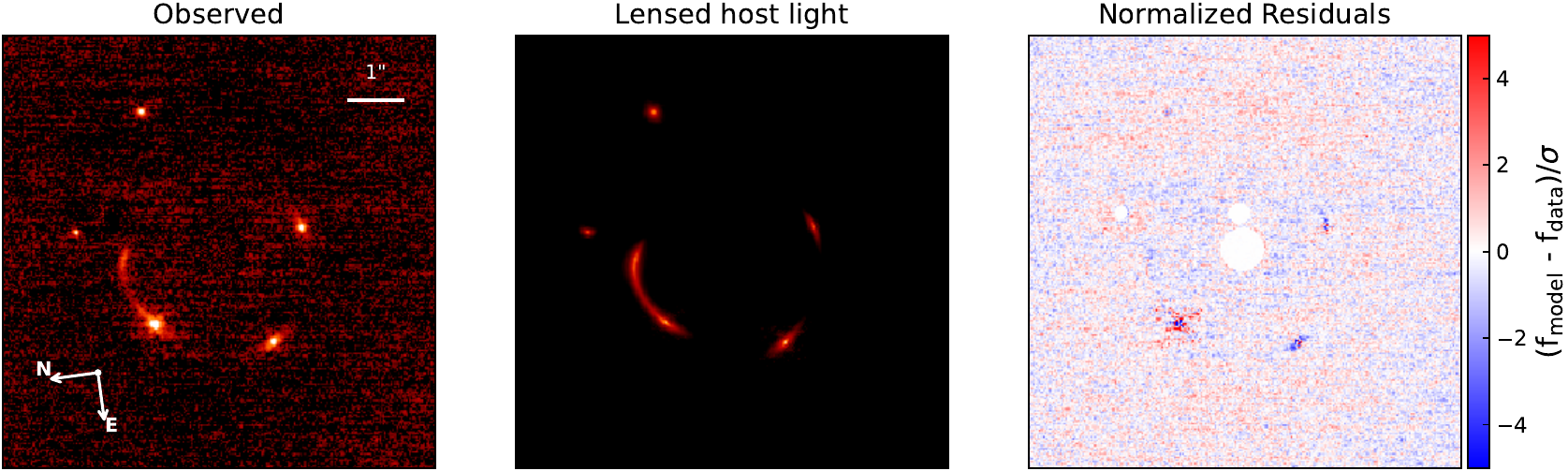}
\caption{
Results of the fiducial lens model fit to the {\it JWST} F115W imaging assuming isothermal density profiles ($\eta_1=\eta_2=2$). The left panel shows the observed image after subtraction of the lens-galaxy light. The middle panel shows the posterior-median reconstruction of the lensed host-galaxy emission obtained from the ensemble of lens-model realizations described in the text. The right panel shows the normalized residuals, $(f_{\rm model}-f_{\rm data})/\sigma$. The white circular regions correspond to masked areas around the centres of lens galaxies G1 and G2 and around the position of source S6, which is not included in the lens modelling. The absence of significant residual structure along the host-galaxy arc indicates that the model provides a good description of the observed surface-brightness distribution.
\label{fig:Modelplot_F115W_eta=2}}
\end{figure*}

To reconstruct the lensed host galaxy while propagating uncertainties in the lens model, we generated an ensemble of host-galaxy realizations by drawing samples from the posterior distributions of the lens and source parameters. For each sample, an image-plane reconstruction was produced using \texttt{lenstronomy}, with the linear surface-brightness amplitudes re-optimized while keeping the sampled nonlinear parameters fixed. We then computed the median surface brightness in each image pixel across the ensemble, yielding a posterior-median reconstruction of the lensed host galaxy. The pixel-by-pixel standard deviation of the ensemble provides a corresponding uncertainty map that incorporates uncertainties in both the lens and source models. An example reconstruction for the F115W data is shown in Figure~\ref{fig:Modelplot_F115W_eta=2}, demonstrating that the inferred lens model successfully reproduces the observed host-galaxy arc morphology.

To investigate the location of the SN relative to its host galaxy, we reconstructed the unlensed source-plane surface-brightness distribution for an ensemble of posterior samples and computed the corresponding source-plane position of the SN by ray tracing the observed images through the lens model. 

\begin{figure}
\centering
\includegraphics[width=\columnwidth]{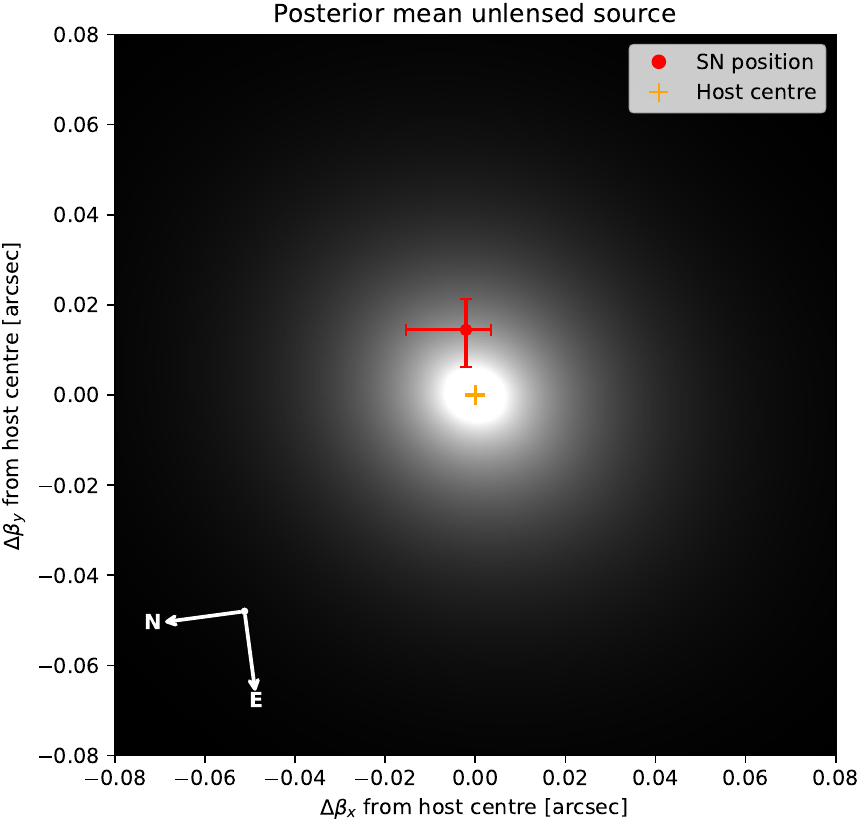}
\caption{
Posterior-mean unlensed host-galaxy reconstruction in the source plane derived from the {\it JWST} F115W imaging assuming isothermal density profiles ($\eta_1=\eta_2=2$). Before combining the posterior samples, each reconstructed host galaxy is recentered on its own light centroid, producing a host-centred reference frame. The red point marks the inferred position of SN~2025wny relative to the centre of its host galaxy, while the error bars show the $68\%$ posterior credible interval of the relative SN--host offset after marginalizing over the lens-model uncertainties. The orange cross marks the host centre, and the compass indicates the orientation on the sky. The SN is located close to the centre of the host galaxy, with a projected offset of approximately $0.14\,\mathrm{kpc}$, corresponding to $\sim0.4\,R_{\rm e}$, where $R_{\rm e}$ is the effective S\'ersic radius of the host galaxy.
\label{fig:Hostplot_F115W_gamma=2_hostcentred}}
\end{figure}

Figure~\ref{fig:Hostplot_F115W_gamma=2_hostcentred} shows the posterior-mean unlensed host-galaxy reconstruction for the F115W data together with the inferred position of SN~2025wny relative to the centre of the host galaxy. To account for correlations between the reconstructed source-plane positions of the SN and the host that arise from the lens-model uncertainties, the host galaxy is recentered for each posterior sample before stacking, and the SN position is shown relative to the corresponding host centre. The plotted uncertainties therefore reflect the posterior distribution of the SN position within its host rather than the uncertainty in the absolute source-plane coordinates.

We find a projected angular separation between the SN and the centre of the host galaxy of
\[
\Delta\theta =
0.016^{+0.012}_{-0.007}\arcsec,
\]
corresponding to a projected physical offset of
\[
\Delta r =
0.14^{+0.10}_{-0.06}\ {\rm kpc}
\]
at the source redshift. The host galaxy is compact, with an effective (S\'ersic) radius of
\[
R_{\rm e} =
0.31^{+0.15}_{-0.07}\ {\rm kpc},
\]
implying a normalized SN offset of
\[
\frac{\Delta r}{R_{\rm e}} =
0.43^{+0.32}_{-0.17},
\]
placing the SN well within the effective radius of its host galaxy.

\begin{table*}[htbp]
\centering
\caption{Constraints on the principal lens-model parameters for the fiducial isothermal model ($\eta_1=\eta_2=2$). Here, quoted uncertainties correspond to $95\,\%$ credible intervals.}
\label{tab:lens_mass}
\begin{tabular}{lccccc}
\toprule
Data & $\theta_{\rm E,1}$ [arcsec] & $q_1$ & $\theta_{\rm E,2}$ [arcsec] & $q_2$ & $\gamma_{\rm ext}$\\
\midrule
F475W & $1.63^{+0.04}_{-0.05}$ & $0.69^{+0.07}_{-0.05}$ & $0.69^{+0.04}_{-0.03}$ & $0.88^{+0.09}_{-0.22}$ & $0.13^{+0.02}_{-0.03}$\\
F625W & $1.58^{+0.11}_{-0.03}$ & $0.73^{+0.03}_{-0.15}$ & $0.75^{+0.02}_{-0.07}$ & $0.86^{+0.10}_{-0.31}$ & $0.11^{+0.02}_{-0.03}$\\
F814W & $1.63^{+0.04}_{-0.04}$ & $0.78^{+0.03}_{-0.18}$ & $0.69^{+0.05}_{-0.03}$ & $0.76^{+0.14}_{-0.30}$ & $0.15^{+0.03}_{-0.03}$\\
F115W & $1.59^{+0.02}_{-0.01}$ & $0.76^{+0.03}_{-0.01}$ & $0.76^{+0.01}_{-0.05}$ & $0.77^{+0.10}_{-0.03}$ & $0.10^{+0.03}_{-0.01}$\\
F150W & $1.59^{+0.02}_{-0.02}$ & $0.73^{+0.03}_{-0.04}$ & $0.76^{+0.01}_{-0.03}$ & $0.78^{+0.17}_{-0.09}$ & $0.11^{+0.01}_{-0.02}$\\
\bottomrule
\end{tabular}
\end{table*}

\begin{table*}[htbp]
\centering
\caption{Predicted image magnifications from the fiducial isothermal lens model ($\eta_1=\eta_2=2$). Negative magnifications correspond to saddle-point images, while positive magnifications correspond to minima of the arrival-time surface. The total magnification, $\mu_{\rm tot}$, is computed from the sum of the absolute image magnifications.}
\label{tab:magnification}
\begin{tabular}{lcccccc}
\toprule
Data & $\mu_{\rm A}$ & $\mu_{\rm B}$ & $\mu_{\rm C}$ & $\mu_{\rm D}$ & $\mu_{\rm E}$ & $\mu_{\rm tot}$\\
\midrule
F475W & $-8.6^{+0.8}_{-1.0}$ & $6.6^{+0.5}_{-0.4}$ & $-3.7^{+0.2}_{-0.2}$ & $3.3^{+0.3}_{-0.2}$ & $-1.8^{+0.2}_{-0.1}$ & $23.9^{+1.7}_{-1.2}$\\
F625W & $-9.7^{+1.0}_{-1.1}$ & $6.4^{+0.3}_{-0.3}$ & $-3.3^{+0.2}_{-0.2}$ & $3.3^{+0.3}_{-0.2}$ & $-1.5^{+0.2}_{-0.1}$ & $24.3^{+1.6}_{-1.8}$\\
F814W & $-13.8^{+1.9}_{-1.9}$ & $7.6^{+0.4}_{-0.4}$ & $-3.9^{+0.2}_{-0.2}$ & $3.8^{+0.3}_{-0.3}$ & $-1.9^{+0.2}_{-0.2}$ & $30.9^{+2.9}_{-2.3}$\\
F115W & $-9.4^{+0.3}_{-0.9}$ & $6.3^{+0.1}_{-0.1}$ & $-3.4^{+0.1}_{-0.1}$ & $4.1^{+0.1}_{-0.1}$ & $-1.8^{+0.1}_{-0.1}$ & $24.9^{+1.1}_{-0.5}$\\
F150W & $-8.3^{+0.3}_{-0.4}$ & $6.2^{+0.1}_{-0.2}$ & $-3.6^{+0.2}_{-0.1}$ & $4.0^{+0.3}_{-0.7}$ & $-1.7^{+0.2}_{-0.1}$ & $23.9^{+0.6}_{-1.5}$\\
\bottomrule
\end{tabular}
\end{table*}

\begin{table*}[htbp]
\centering
\caption{Predicted image magnifications after incorporating microlensing for the fiducial isothermal lens model ($\eta_1=\eta_2=2$). Quoted uncertainties include contributions from both the macrolens model and microlensing.}
\label{tab:magnification_microlensing}
\begin{tabular}{lcccccc}
\toprule
Data & $\mu_{\rm A}$ & $\mu_{\rm B}$ & $\mu_{\rm C}$ & $\mu_{\rm D}$ & $\mu_{\rm E}$ & $\mu_{\rm tot}$\\
\midrule
F475W & $-4.9^{+3.0}_{-12.9}$ & $4.9^{+5.8}_{-1.3}$ & $-2.0^{+1.0}_{-4.0}$ & $3.0^{+2.1}_{-0.4}$ & $-1.5^{+0.6}_{-1.0}$ & $13.1^{+13.2}_{-6.4}$\\
F625W & $-5.4^{+3.2}_{-14.6}$ & $4.6^{+5.5}_{-1.2}$ & $-1.8^{+0.9}_{-3.6}$ & $3.0^{+2.2}_{-0.4}$ & $-1.2^{+0.5}_{-0.8}$ & $13.4^{+14.5}_{-6.6}$\\
F814W & $-7.5^{+4.6}_{-20.5}$ & $5.5^{+6.5}_{-1.4}$ & $-2.1^{+1.1}_{-4.2}$ & $3.4^{+2.6}_{-0.5}$ & $-1.6^{+0.6}_{-1.1}$ & $17.2^{+20.1}_{-8.5}$\\
F115W & $-4.7^{+2.6}_{-13.3}$ & $4.4^{+5.0}_{-0.9}$ & $-1.9^{+1.0}_{-3.5}$ & $3.2^{+1.2}_{-0.3}$ & $-1.5^{+0.6}_{-0.9}$ & $12.2^{+13.2}_{-5.6}$\\
F150W & $-4.0^{+2.2}_{-11.5}$ & $4.3^{+4.9}_{-0.9}$ & $-2.0^{+1.1}_{-3.7}$ & $3.1^{+1.1}_{-0.5}$ & $-1.4^{+0.5}_{-0.9}$ & $11.3^{+11.7}_{-5.2}$\\
\bottomrule
\end{tabular}
\end{table*}

\begin{table*}[htbp]
\centering
\caption{Observed and modelled flux ratios relative to image A. The observed ratios are given by $f_{\rm X}/f_{\rm A}$, and the modelled ratios by $\mu_{\rm X}/\mu_{\rm A}$. The model predictions are based on the fiducial isothermal lens model ($\eta_1=\eta_2=2$) and include the effects of microlensing. The observed flux ratios do not account for the phase differences between the SN images, which are expected to be most important for image D owing to its relatively large time delay.}
\label{tab:ratios}
\begin{tabular}{lcccccc}
\toprule
Image & $f_{\rm X}/f_{\rm A}$ (F475W) & $f_{\rm X}/f_{\rm A}$ (F625W) & $f_{\rm X}/f_{\rm A}$ (F814W) & $f_{\rm X}/f_{\rm A}$ (F115W) & $f_{\rm X}/f_{\rm A}$ (F150W) & $\mu_{\rm X}/\mu_{\rm A}$ (modelled)\\
\midrule
B & $0.271 \pm 0.004$ & $0.241 \pm 0.003$ & $0.231 \pm 0.003$ & $0.242 \pm 0.003$ & $0.249 \pm 0.003$ & $1.21^{+2.01}_{-0.91}$\\
C & $0.237 \pm 0.004$ & $0.210 \pm 0.003$ & $0.201 \pm 0.003$ & $0.239 \pm 0.003$ & $0.216 \pm 0.003$ & $0.46^{+1.27}_{-0.35}$\\
D & $0.159 \pm 0.003$ & $0.098 \pm 0.002$ & $0.088 \pm 0.002$ & $0.086 \pm 0.002$ & $0.093 \pm 0.002$ & $0.82^{+1.10}_{-0.61}$\\
E & $0.054 \pm 0.001$ & $0.045 \pm 0.002$ & $0.045 \pm 0.002$ & $0.057 \pm 0.002$ & $0.046 \pm 0.002$ & $0.32^{+0.56}_{-0.24}$\\
\bottomrule
\end{tabular}
\end{table*}

%-----------------------------------------------------------------------
\subsection{Time-delay cosmography}
%-----------------------------------------------------------------------

The time delays between the lensed images of SN~2025wny provide a direct probe of the Hubble constant through the time-delay distance \citep{refsdal_1964b}. For a fixed lens mass model, the predicted time delays scale inversely with the Hubble constant,
\begin{equation}
\Delta t \propto H_0^{-1}.
\end{equation}
Consequently, a comparison between the model-predicted time delays and the observed delays can be used to infer $H_0$.

For each sample of the lens-model posterior, we compute the arrival times of all predicted images assuming a reference cosmology with $H_0=70\;{\rm km\,s^{-1}\,Mpc^{-1}}$. This yields model predictions for the relative time delays, $\Delta t_{70}$, between each image and the reference image A. Given an observed time delay $\Delta t_{\rm obs}$, the corresponding value of the Hubble constant is then
\begin{equation}
H_0 =
70\,
\frac{\Delta t_{70}}
{\Delta t_{\rm obs}}
\ {\rm km\,s^{-1}\,Mpc^{-1}}.
\label{eq:H0_scaling}
\end{equation}

%-----------------------------------------------------------------------
\subsection{Observed time delays}
%-----------------------------------------------------------------------

The relative time delays between the lensed images of SN~2025wny were measured by \companioncitet{Johansson2026wny} using time-resolved spectroscopy. The analysis exploits the temporal evolution of broad emission and absorption features in the SN spectrum, tracking their wavelength positions as a function of phase for each image. These features provide a robust tracer of the spectroscopic evolution and are less sensitive to absolute flux calibration and microlensing-induced magnification than continuum-based methods.

Independent measurements of the $\Delta t_{AB}$ and $\Delta t_{AC}$ time delays were obtained by \companioncitet{Townsend2026wny} from multi-band photometric monitoring. Their analysis combines scene-modelling photometry with Gaussian-process modelling of the SN light curves to infer the relative arrival times of the resolved images while simultaneously accounting for their relative magnifications. These measurements are statistically consistent with the spectroscopic results and are combined below using inverse-variance weighting to obtain the adopted constraints on $\Delta t_{AB}$ and $\Delta t_{AC}$.

The inferred time delays relative to image A are summarized in Table~\ref{tab:observed_delays}. Throughout this work, we adopt the convention
\begin{equation}
\Delta t_{AX} \equiv t_X - t_A,
\end{equation}
such that negative values indicate images arriving before image A. For $\Delta t_{AB}$ and $\Delta t_{AC}$, we also present combined constraints obtained as inverse-variance weighted averages of the independent spectroscopic and photometric measurements. Since the two methods are statistically consistent, no additional uncertainty inflation is applied.
The quoted uncertainties correspond to $68\,\%$ credible intervals. 

\begin{table}[htbp]
\centering
\caption{Observed time delays of SN~2025wny relative to image A. Spectroscopic measurements are from \companioncitet{Johansson2026wny}, photometric measurements from \companioncitet{Townsend2026wny}. For $\Delta t_{AB}$ and $\Delta t_{AC}$, we also list the inverse-variance weighted combination of the two independent measurements.}
\label{tab:observed_delays}
\begin{tabular}{lccc}
\toprule
Delay & Spectroscopic & Photometric & Combined \\
 & [days] & [days] & [days] \\
\midrule
$\Delta t_{AB}$ & $-10.3 \pm 2.3$ & $-10.6 \pm 2.4$ & $-10.4 \pm 1.7$ \\
$\Delta t_{AC}$ & $0.1 \pm 3.6$ & $1.2 \pm 2.6$ & $0.8 \pm 2.1$ \\
$\Delta t_{AD}$ & $-65.7 \pm 3.5$ & --- & --- \\
$\Delta t_{AE}$ & $3.7 \pm 8.8$ & --- & --- \\
\bottomrule
\end{tabular}
\end{table}

%-----------------------------------------------------------------------
\subsection{Hubble constant constraints}
%-----------------------------------------------------------------------

For each posterior sample of the lens model, we combine the predicted time delays with random realizations of the observed delays---using the combined spectroscopic and photometric constraints for $\Delta t_{AB}$ and $\Delta t_{AC}$, and the spectroscopic constraints for $\Delta t_{AD}$ and $\Delta t_{AE}$---and infer the corresponding value of $H_0$ using Equation~(\ref{eq:H0_scaling}). In this way, both the uncertainties in the lens model and those of the observed time delays are propagated through a Monte Carlo analysis, yielding the posterior distributions and credible intervals quoted below.

We first perform this inference independently for each filter. Since all filters are compared with the same observed time delays, the resulting $H_0$ posteriors are correlated and cannot be regarded as statistically independent measurements. We therefore treat the choice of filter as a modelling uncertainty and construct a filter-marginalized posterior by combining samples drawn with equal weight from the individual filter posteriors. This procedure incorporates both the statistical uncertainty within each filter and the systematic scatter associated with the imaging data used for the lens modelling.
The model-predicted and observed time delays were determined
independently. The results presented below therefore represent a blind combination of
the two analyses.

For the fiducial model, in which the density slopes of both lens galaxies are fixed to $\eta_1=\eta_2=2$, the inferred values of $H_0$ are listed in Table~\ref{tab:H0}. Combining the F475W, F625W, F814W, F115W, and F150W results gives
\[
H_0 = 66.7^{+7.6}_{-6.3}\; {\rm km\,s^{-1}\,Mpc^{-1}}.
\]
This estimate is conservative in the sense that the filter-to-filter scatter, $\sigma(H_0)\simeq 4.6\,{\rm km\,s^{-1}\,Mpc^{-1}}$, is treated as a systematic modelling uncertainty. Restricting the analysis to the two highest-resolution {\it JWST} filters with the clearest host-galaxy arc detections, F115W and F150W, gives
\[
H_0 = 66.1^{+6.4}_{-4.5}\; {\rm km\,s^{-1}\,Mpc^{-1}}.
\]

\begin{figure*}[htbp]
    \centering
    \includegraphics[width=0.6\linewidth]{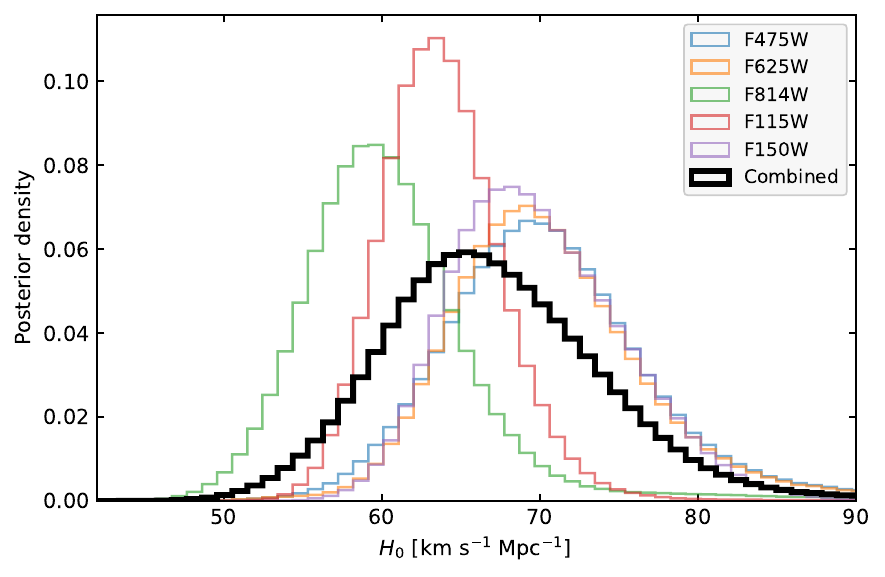}		
    \caption{Posterior distributions of the Hubble constant obtained from the individual lens models assuming isothermal density profiles ($\eta_1=\eta_2=2$). The coloured histograms show the results derived from the five filters independently, while the black histogram shows the filter-marginalized posterior obtained by combining all filters with equal weight. The individual constraints are broadly consistent, indicating that the inferred value of $H_0$ is relatively insensitive to the choice of filter used in the lens modelling. The width of the combined posterior incorporates both the statistical uncertainty within each filter and the systematic uncertainty associated with filter-dependent modelling differences.
\label{fig:H0_eta=2_comb}}
\end{figure*}

Figure~\ref{fig:H0_eta=2_comb} shows the posterior distributions of $H_0$ obtained from the individual filters and from the combined analysis. Although the posterior medians span the range $H_0 \simeq 60\text{--}70\;{\rm km\,s^{-1}\,Mpc^{-1}}$, the individual constraints exhibit substantial overlap, indicating that filter-dependent modelling uncertainties are smaller than the statistical uncertainty of the measurement. The combined posterior therefore provides a robust estimate of $H_0$ that incorporates both statistical uncertainties and the systematic uncertainty associated with the choice of filter used in the lens modelling.

\begin{figure*}[htbp]
    \centering
    \includegraphics[width=0.7\linewidth]{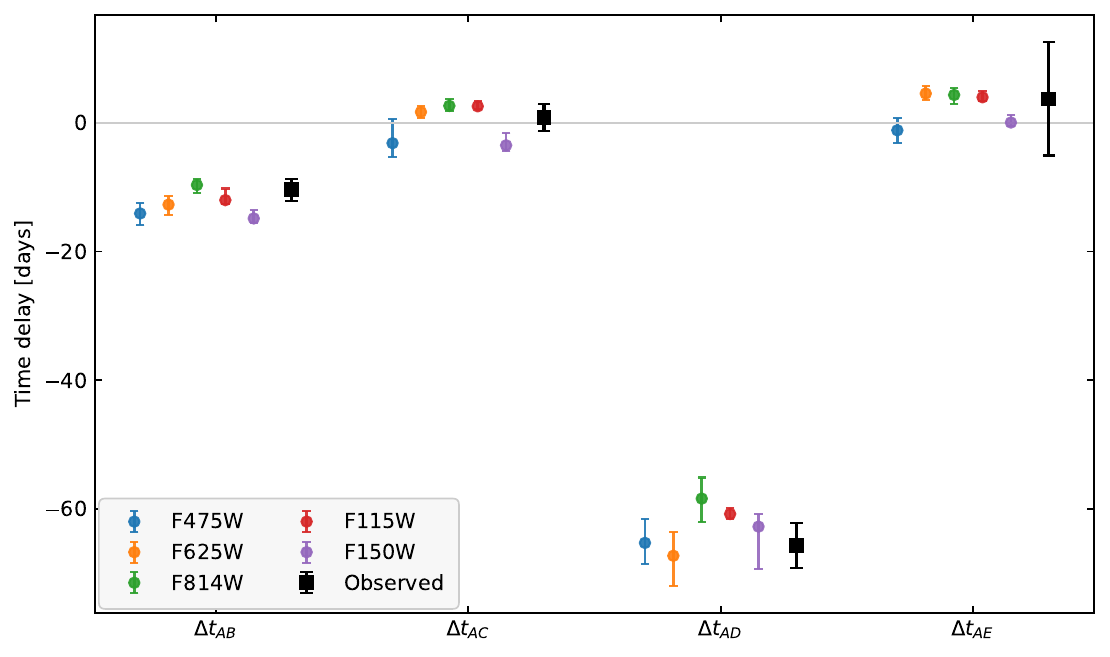}		
    \caption{Comparison of the observed and model-predicted time delays for the fiducial isothermal lens model ($\eta_1=\eta_2=2$). Colored points show the median predicted time delays for the individual filters after rescaling to the best-fitting combined value of $H_0=66.7\,{\rm km\,s^{-1}\,Mpc^{-1}}$, with error bars indicating the central $68\,\%$ credible intervals.     Black points show the adopted observed time delays, using the combined spectroscopic and photometric constraints for $\Delta t_{AB}$ and $\Delta t_{AC}$ and the spectroscopic constraints for $\Delta t_{AD}$
    and $\Delta t_{AE}$.
    The agreement between the model predictions and observations is generally good across all image pairs. The longest delay, $\Delta t_{AD}$, provides the strongest constraint on $H_0$ owing to its relatively small fractional uncertainty.
\label{fig:Time_delays_eta=2_comb}}
\end{figure*}

Figure~\ref{fig:Time_delays_eta=2_comb} compares the observed and model-predicted time delays for the fiducial model, with the model delays rescaled to the best-fitting combined value of $H_0=66.7\;{\rm km\,s^{-1}\,Mpc^{-1}}$. The agreement between the predicted and observed delays is generally good across all image pairs. Because the uncertainty on $H_0$ is determined primarily by the fractional precision of the time-delay measurements, the longest measured delay, $\Delta t_{AD}$, provides the strongest individual constraint. Using only $\Delta t_{AD}$ yields
\[
H_0 = 64.0^{+6.9}_{-5.3}\;{\rm km\,s^{-1}\,Mpc^{-1}},
\]
close to the result obtained from the full set of time delays. In contrast, using only $\Delta t_{AB}$ gives
\[
H_0 = 81.4^{+23.3}_{-18.7}\;{\rm km\,s^{-1}\,Mpc^{-1}},
\]
demonstrating that the shorter-delay image pairs contribute comparatively little additional constraining power.

To quantify the impact of the assumed mass-profile slopes, we also fit models to the F115W and F150W data in which the density slopes $\eta_1$ and $\eta_2$ are allowed to vary. The resulting slope constraints are listed in Table~\ref{tab:lens_mass_slope}. The inferred density slopes are consistent between the two {\it JWST} filters. The primary lens galaxy, G1, is consistent with an approximately isothermal profile, with $\eta_1 \simeq 1.9$, while the secondary lens galaxy, G2, prefers a steeper density profile with $\eta_2 \simeq 2.3$. 
We note that the posterior distribution for the slope of the secondary lens galaxy approaches the upper boundary of the adopted prior, indicating that this parameter is not fully constrained by the present data.
The agreement between the F115W and F150W results suggests that the slope constraints are driven primarily by the common host-galaxy morphology rather than by filter-dependent modelling systematics.

For these free-slope models, we obtain
\[
H_0 = 70.8^{+8.2}_{-6.1}\; {\rm km\,s^{-1}\,Mpc^{-1}}.
\]
Allowing the density slopes to vary shifts the inferred value of the Hubble constant upward by $\sim 4\;{\rm km\,s^{-1}\,Mpc^{-1}}$ relative to the fiducial isothermal model, while bringing the independent F115W and F150W constraints into closer agreement. The increase in $H_0$ remains smaller than the statistical uncertainty of the measurement. However, because this comparison uses only the two {\it JWST} filters, it provides a less conservative estimate of filter-dependent modelling systematics than the full five-filter analysis.

\begin{table}[htbp]
\centering
\caption{Constraints on the Hubble constant derived from the observed time delays and the lens models described in the text. Quoted uncertainties correspond to the central $68\,\%$ credible intervals. Values of $H_0$ are given in units of ${\rm km\,s^{-1}\,Mpc^{-1}}$.}
\label{tab:H0}
\begin{tabular}{llc}
\toprule
Model & Filter & $H_0$ \\
\midrule
$\eta=2$ & F475W & $70.2^{+6.8}_{-5.8}$ \\
         & F625W & $70.3^{+7.4}_{-5.3}$ \\
         & F814W & $60.1^{+5.6}_{-4.4}$ \\
         & F115W & $63.6^{+3.8}_{-3.4}$ \\
         & F150W & $69.4^{+6.0}_{-4.8}$ \\ \hline
         & Combined & $66.7^{+7.6}_{-6.3}$ \\ 
\midrule \midrule 
$\eta$ free & F115W & $71.6^{+10.2}_{-6.4}$ \\
            & F150W & $70.1^{+6.6}_{-5.9}$ \\ \hline 
            & Combined & $70.8^{+8.2}_{-6.1}$ \\
\bottomrule
\end{tabular}
\end{table}

\begin{table}[htbp]
\centering
\caption{Constraints on the density slopes of the two lens galaxies obtained from the free-slope lens models. Uniform priors were adopted over the interval $1.6 \leq \eta_i \leq 2.4$.}
\label{tab:lens_mass_slope}
\begin{tabular}{lcc}
\toprule
Filter & $\eta_1$ & $\eta_2$ \\
\midrule
F115W & $1.94^{+0.19}_{-0.12}$ & $2.33^{+0.06}_{-0.08}$ \\
F150W & $1.86^{+0.13}_{-0.11}$ & $2.30^{+0.09}_{-0.19}$ \\
\bottomrule
\end{tabular}
\end{table}

%-----------------------------------------------------------------------
\subsection{Impact of the mass-sheet degeneracy}\label{appsec:MSD}
%-----------------------------------------------------------------------

Strong gravitational lensing is subject to the mass-sheet degeneracy
(MSD), whereby a transformation of the projected mass distribution,
\begin{equation}
\kappa(\boldsymbol{\theta})
\rightarrow
\xi\,\kappa(\boldsymbol{\theta}) + (1-\xi),
\end{equation}
leaves the observed image positions and shapes unchanged while
rescaling the source plane \citep[e.g.,][]{Falco1985}. 
Since modelled time
delays scale as  
\begin{equation}
\Delta t(\xi)=\xi\,\Delta t(\xi=1),
\end{equation}
the inferred Hubble constant transforms
according to
\begin{equation}
H_0(\xi)=\xi\,H_0(\xi=1),
\end{equation}
such that time-delay cosmography constrains a degeneracy track in the
$(H_0,\xi)$ plane rather than a unique value of $H_0$.

The MSD may be particularly relevant for SN~2025wny since the lens system
appears to reside in an overdense environment. The system was
independently identified as a galaxy cluster candidate by
\citet{Wen2024}, who searched for overdensities in the projected
stellar-mass distribution of galaxies in the DESI Legacy Imaging
Surveys. Assuming the primary lens galaxy G1 to be the brightest
cluster galaxy, \citet{Wen2024} identified 16 candidate cluster members
and inferred a cluster mass of $M_{500}\approx 1.45\cdot 10^{14}\,M_\odot$.
Here, $M_{500}$ denotes the mass enclosed within the radius $R_{500}$
inside which the mean density equals
$500\,\rho_{\rm crit}(z)$. Such a cluster environment may contribute a
non-negligible external convergence, thereby shifting the inferred
value of $H_0$. A detailed analysis of the mass distribution in the
field of SN~2025wny will be presented in a future paper.

\begin{figure*}[htbp]
    \centering
    \includegraphics[width=0.6\linewidth]{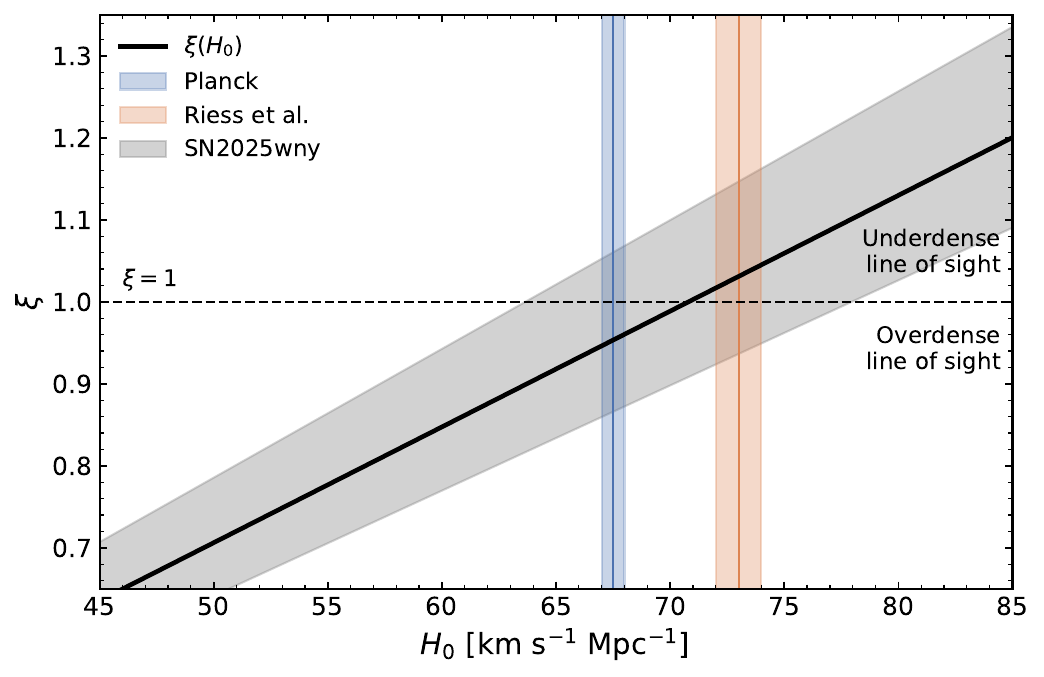}
    \caption{
Constraint in the $(H_0,\xi)$ plane implied by the combined free-slope
F115W and F150W lens models of SN~2025wny. The solid curve shows the
median mass-sheet degeneracy relation, while the shaded region
indicates the corresponding $68\,\%$ credible interval. The horizontal
dashed line marks the fiducial assumption $\xi=1$ adopted throughout
this work, corresponding to no additional mass-sheet transformation.
The vertical shaded bands show the $1\sigma$ constraints from
\textit{Planck} 2018 under flat $\Lambda$CDM \citep{planck18} and from
the local distance-ladder determination of
\citet{2022ApJ...934L...7R}. Their intersections with the lensing
constraint imply values of $\xi$ close to unity, suggesting that large
corrections from the mass-sheet degeneracy are disfavoured.
    \label{fig:H0_xi_lensing_constraint_comb}}
\end{figure*}

Figure~\ref{fig:H0_xi_lensing_constraint_comb} shows the resulting
constraint in the $(H_0,\xi)$ plane derived from the combined free-slope F115W
and F150W lens models. Conversely, external measurements of the Hubble
constant can be used to constrain the magnitude of the mass-sheet
transformation. Adopting the Planck 2018 value,
$H_0 = 67.5 \pm 0.5\;{\rm km\,s^{-1}\,Mpc^{-1}}$
\citep{planck18}, we obtain
\[
\xi = 0.95 \pm 0.10.
\]
Using the local distance-ladder measurement of
\citet{2022ApJ...934L...7R},
$H_0 = 73.0 \pm 1.0\;{\rm km\,s^{-1}\,Mpc^{-1}},$
gives
\[
\xi = 1.03 \pm 0.10.
\]

Both estimates are consistent with $\xi=1$, suggesting
that any external convergence contribution is modest. Interpreted in
terms of a uniform mass sheet, this corresponds to
\[
\kappa_{\rm ext}=1-\xi,
\]
implying little room for a large positive convergence from the
surrounding environment.

The MSD also affects the inferred image magnifications. Under a
mass-sheet transformation, the total magnification scales as
\[
\mu_{\rm tot}(\xi)=\frac{\mu_{\rm tot}(\xi=1)}{\xi^2},
\]
where $\mu_{\rm tot}(\xi=1)$ is the magnification inferred from the fiducial
lens model. Since the preferred values of $\xi$ remain close to unity,
the corresponding corrections to the magnifications are modest.

%-----------------------------------------------------------------------
\subsection{Stellar velocity dispersions}
%-----------------------------------------------------------------------

Stellar kinematics provide an independent constraint on the mass
distribution of the lens galaxies and therefore complement the lensing
constraints. In particular, they can help constrain the mass-sheet
degeneracy discussed in the previous section, since the stellar velocity
dispersion probes the gravitational potential of the lens galaxy rather
than the additional uniform mass sheet introduced by the mass-sheet
transformation. Under the transformation introduced above, one expects
$\sigma_v\propto\sqrt{\xi}$ \citep{SchneiderSluse2013},
such that a positive external convergence lowers the velocity
dispersion predicted for the lens galaxy while leaving the lensing
observables largely unchanged.

As an independent consistency check, we compare the stellar velocity
dispersions predicted from our lens models with the spectroscopically
inferred dispersions of G1 and G2 (see \companioncitet{Johansson2026wny} for
details). For the free-slope models, the lensing-only solutions
generally predict a lower velocity dispersion for G1 and a higher
velocity dispersion for G2 than inferred from the spectroscopy,
reflecting in particular the preference for a relatively steep mass
profile in G2. Including the spectroscopic velocity dispersions as an
additional likelihood term shifts the preferred mass models toward
better agreement with the kinematic measurements, although some residual
tension remains.

Interpreting this tension is not straightforward, however, since
relating a lens mass model to an observed stellar velocity dispersion
requires several assumptions that are largely independent of the lensing
analysis. We compute the dynamical predictions using the spherical Jeans
approximation \citep{BinneyTremaine2008}, despite the fact that both G1
and G2 are elliptical and form a close pair. We further
assume isotropic stellar orbits ($\beta=0$), although different
anisotropy profiles can produce similar projected velocity dispersions
through the well-known mass--anisotropy degeneracy
\citep{BinneyMamon1982,Tonry1983}.

The predicted velocity dispersions also depend on the adopted stellar
light tracer and effective spectroscopic aperture. We use the F814W
surface-brightness model as the tracer population and evaluate the Jeans
equation within an aperture of twice the measured effective radius.
However, the observed dispersions are inferred from spectra using
many absorption features spanning a broad wavelength range, making it
unclear which photometric band best traces the stellar population
dominating the measured kinematic signal. Likewise, the effective
aperture depends on wavelength dependent seeing conditions and PSFs. Recent work has shown that these effects, together with
uncertainties in the orbital anisotropy and light-profile modelling,
can introduce substantial systematic uncertainties \citep{ForesToribio2026}.

Despite these caveats, the stellar velocity dispersions provide a useful
consistency check on the lens models. Including the kinematic
constraints modifies the preferred mass profiles but has only a modest
effect on the inferred value of $H_0$. For the free-slope models based
on the F115W and F150W data, the inferred Hubble constant increases by
approximately $5\,{\rm km\,s^{-1}\,Mpc^{-1}}$, corresponding to
$\lesssim 0.6\sigma$ of the statistical uncertainty. For our fiducial models
with $\eta_1=\eta_2=2$, the shift is less than
$2\,{\rm km\,s^{-1}\,Mpc^{-1}}$ ($\lesssim 0.3\sigma$), since the stellar
velocity dispersions are then determined primarily by the
well-constrained Einstein radii. Our principal cosmological conclusions
therefore remain robust to the inclusion of the current kinematic
constraints.

Future high signal-to-noise, spatially resolved integral-field
spectroscopy of G1 and G2 could potentially reduce the present
systematic uncertainties by directly constraining the two-dimensional
stellar kinematics, thereby relaxing several of the assumptions
underlying the present Jeans modelling and enabling more
realistic dynamical analyses.

%=======================================================================
\section{Discussion \label{sec:discussion}}
%=======================================================================
%-----------------------------------------------------------------------
\subsection{Implications for the intrinsic luminosity of SN~2025wny}
%-----------------------------------------------------------------------

The lens modelling presented in this work implies that SN~2025wny is
strongly magnified, with a total magnification of order
$\mu_{\rm tot}\sim 5$--$50$ once microlensing is taken into account.
Correcting the observed brightness for this magnification can substantially
reduce the inferred intrinsic luminosity of the event.

Although a detailed analysis of the SN properties is beyond the
scope of this paper and will be presented by 
\companioncitet{Li2026wny}, our
results indicate that the intrinsic luminosity of SN~2025wny is broadly
consistent with the range observed among previously studied SLSN-I. The extreme apparent brightness of the event
may therefore primarily be a consequence of gravitational lensing rather
than requiring an intrinsically exceptional explosion.

Selection effects may further favor the higher-magnification end of the
allowed parameter space. Since strongly magnified transients are more
likely to be discovered and followed up than weakly magnified events,
the observed population of lensed SNe is subject to
magnification bias. Consequently, although the posterior distributions
derived in this work formally extend to relatively modest
magnifications, the fact that SN~2025wny was identified as an unusually
bright transient at $z=2.015$ increases the a priori likelihood that it
resides in the higher-magnification tail of the distribution. A
quantitative treatment of this effect is beyond the scope
of the present work and will be presented in a separate paper. 

%-----------------------------------------------------------------------
\subsection{Additional images}
%-----------------------------------------------------------------------
For each posterior sample of the lens model, we solved the lens equation
to identify all images associated with the inferred source position,
including images that were not used in the fitting procedure. This
provides a consistency check on the inferred lens model and allows us to
assess whether acceptable models predict additional supernova images
beyond the five observed images.

Our methodology differs from that of \citet{2026arXiv260216620E}. In
their analysis, the source position was treated as a free model
parameter, allowing solutions that predict additional images to be
excluded directly during the fitting procedure. In contrast,
\texttt{lenstronomy} parameterizes the observed image positions and
evaluates their consistency in the source plane. The presence or absence
of additional images is therefore not imposed as a prior constraint in
our modelling. Instead, we evaluate the image multiplicity a posteriori
by solving the lens equation for posterior samples drawn from the MCMC chains.

For the fiducial F115W model with $\eta_1=\eta_2=2$, we analysed
$10^4$ posterior realizations. The vast majority ($98.0\,\%$) predict
only the five observed images. However, $1.3\,\%$ of the realizations
predict seven images, corresponding to the creation of an additional
image pair, while $0.7\,\%$ return six images. Since these six-image
solutions do not appear to result from closely merged image pairs, we
interpret them conservatively as numerical or topology-edge cases and do
not use them for the quantitative characterization of the additional
image pair.

\begin{figure}[htbp]
    \centering
    \includegraphics[width=\linewidth]{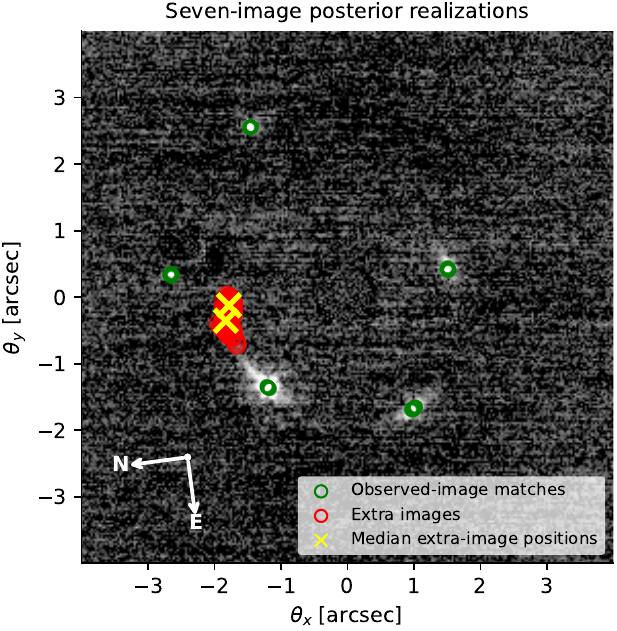}		
    \caption{
    Predicted SN image positions for posterior realizations of the fiducial
    F115W lens model ($\eta_1=\eta_2=2$) that produce seven images. The
    background shows the F115W image after subtraction of the lens
    galaxies. Green circles indicate the five observed SN images, while red
    circles show the additional image pair predicted by these realizations.
    The yellow crosses mark the median positions of the two additional
    images. The additional images are concentrated along the host-galaxy arc
    between images A and E, close to the critical curve, indicating that
    they arise when the source lies close to a caustic crossing.
    \label{fig:Predicted_image_positions_F115W_eta=2}}
\end{figure}

The seven-image solutions predict an additional image pair located along
the host-galaxy arc between images A and E
(Figure~\ref{fig:Predicted_image_positions_F115W_eta=2}), close to the
critical curve. The predicted positions are tightly clustered,
indicating that the image pair is associated with a localized caustic
crossing. For these realizations, the two additional images have median
magnifications of
\[
|\mu| = 13.5^{+9.1}_{-3.1}
\]
and
\[
|\mu| = 23.5^{+5.6}_{-6.6},
\]
respectively, where the uncertainties correspond to the
$16$th and $84$th percentiles of the posterior distribution. The
corresponding time delays relative to image A are
\[
\Delta t = -1.3\pm 0.6\;{\rm days}
\]
and
\[
\Delta t = -1.4^{+0.7}_{-0.5}\ {\rm days},
\]
respectively.

Although only a small fraction ($1.3\,\%$) of the posterior samples predict an additional image pair, the inferred magnifications imply that these images would be brighter than several of the observed SN images and should therefore have been readily detected in the available imaging. Their absence thus provides an independent observational argument against these realizations. We nevertheless retain them in the posterior analysis, since the presence or absence of additional images is not included as an explicit likelihood constraint in our \texttt{lenstronomy} modelling. Given the small fraction of affected posterior samples, excluding these realizations would be unlikely to have a significant impact on the inferred lens-model parameters or the derived value of $H_0$. Their occurrence nevertheless demonstrates that the source lies close to a caustic, such that relatively small changes in the lens parameters can alter the image multiplicity.

%-----------------------------------------------------------------------
\subsection{SN~2025wny as a cosmographic probe}
%-----------------------------------------------------------------------

SN~2025wny occupies a unique region of parameter space among currently known lensed supernovae. In addition to being the first known strongly lensed Type~I superluminous supernova, it is among the highest-redshift lensed supernovae for which the multiple images can be individually resolved and monitored from the ground. 
The combination of high magnification, multiple resolved images, and measurable spectroscopic time delays demonstrates the potential of lensed supernovae as independent probes of cosmology. Future wide-field
time-domain surveys are expected to increase the sample of such systems substantially, enabling both more precise measurements of $H_0$ and improved constraints on the mass distributions of lens galaxies.

The cosmographic constraints presented here remain conditional on the adopted parameterization of the lens mass distribution. Although we have explored the impact of allowing the density slopes of the lens
galaxies to vary, and find that incorporating the currently available stellar-kinematic measurements has only a modest effect on $H_0$, the dynamical analysis remains subject to uncertainties associated with the stellar light distribution, orbital anisotropy, and effective
spectroscopic aperture. Improved kinematic measurements will therefore provide an important consistency check on the lens models and help
reduce systematic uncertainties in the inferred value of $H_0$.

The inferred values of $H_0$ are furthermore subject to the mass-sheet degeneracy. Throughout this work we have assumed $\xi=1$, corresponding to the absence of an additional mass-sheet transformation. The apparent association of SN~2025wny with a galaxy overdensity makes this effect particularly relevant, since an external mass distribution may contribute a non-zero convergence, $\kappa_{\rm ext}$. However, the magnitude of this correction depends on the detailed distribution of matter along the line of sight and cannot be inferred reliably from the presence of a galaxy overdensity alone.

A robust assessment of the environmental contribution will require a dedicated analysis of the surrounding mass distribution using spectroscopic and photometric observations and will be presented in future work. Together with independent stellar-kinematic constraints on the lens galaxies, such measurements will enable a more robust cosmographic determination of $H_0$ from SN~2025wny.

%=======================================================================
\section{Conclusions \label{sec:conclusion}}
%=======================================================================

We have presented a lensing and cosmographic analysis of the strongly
lensed Type~I superluminous supernova SN~2025wny using imaging from
{\it HST} and {\it JWST}. Our main conclusions are as follows:

\begin{enumerate}

\item We model the lens system using two power-law mass distributions,
representing the two lens galaxies, together with an external shear
component. The inferred Einstein radii are approximately
$\theta_{\rm E,1}\simeq1.6\arcsec$ and
$\theta_{\rm E,2}\simeq0.7$--$0.8\arcsec$, in good agreement across all
filters and broadly consistent with previous modelling based on
ground-based adaptive-optics imaging.

\item The lens models imply substantial magnification of the SN.
Including microlensing by stars in the lens galaxies broadens the
magnification distributions significantly and shifts the posterior
medians toward lower values. We infer a total magnification of
$\mu_{\rm tot}\sim5$--$50$ ($95\,\%$ credible intervals), depending on the adopted filter.

\item The observed flux ratios of the multiple images are broadly
consistent with the combined macro- and microlensing predictions.
Although some image pairs show mild tension with the posterior medians,
all measured flux ratios are consistent with the predicted
$95\,\%$ credible intervals.

\item Combining the measured spectroscopic and photometric time delays with the lens
models yields a filter-marginalized constraint of
\[
H_0 = 66.7^{+7.6}_{-6.3}\;
{\rm km\,s^{-1}\,Mpc^{-1}}
\]
for the fiducial isothermal model. Allowing the density slopes of the
lens galaxies to vary gives
\[
H_0 = 70.8^{+8.2}_{-6.1}\;
{\rm km\,s^{-1}\,Mpc^{-1}},
\]
showing that allowing the power-law density slopes to vary within the
adopted priors produces a shift smaller than the current statistical
uncertainty.

\item The longest measured delay, $\Delta t_{AD}$, provides the dominant
contribution to the cosmological constraint. The consistency of the
results obtained from independent filters indicates that filter-dependent
modelling systematics are subdominant to the current statistical
uncertainties.

\item We investigate the impact of the mass-sheet degeneracy and emphasize
that the inferred values of $H_0$ remain conditional on assumptions
about both the internal mass profile of the lens galaxies and any
additional convergence from the surrounding environment. Future
measurements of the stellar velocity dispersions of the lens galaxies,
together with a detailed characterization of the lens environment, will
be required for a fully robust cosmographic determination of $H_0$.

\end{enumerate}

SN~2025wny is the first known strongly lensed superluminous SN and provides a valuable addition to the small sample of lensed supernovae suitable for cosmographic studies. The combination of high-resolution {\it HST} and {\it JWST} imaging with spectroscopically measured time delays demonstrates the potential of this system for precision cosmography. Future measurements of the lens-galaxy kinematics and a detailed characterization of the lens environment will further strengthen its utility as a cosmological probe.

\normalsize
\vspace{1.5cm}
%\section*{Data Availability}

\section*{Acknowledgments}
\input{acknowledgments}

\bibliography{references}   
%\bibliography{references, Lensbib}   

\appendix{}

%=======================================================================
\section{Microlensing modelling}\label{appsec:microlensingmodelling}
%=======================================================================

Microlensing magnification probability distributions are generated using the microlensing simulator of \url{https://gloton.ugr.es/microlensing/}. The simulations require the local convergence $\kappa$, local shear amplitude $|\gamma|$, and stellar mass fraction $f_\star$ at each image position. 
We estimate the stellar mass fraction at the lensed image positions by combining the macrolens model with the observed light distribution of the lens galaxies.

The stellar masses of the two main lens galaxies are estimated within circular apertures of radius $1.0\arcsec$ centred on each galaxy. The inferred masses are \companioncitep{Goobar2026wny}
\[
\log (M_\star/M_\odot) = 11.32^{+0.13}_{-0.17} \quad (\mathrm{G1}),
\qquad
\log (M_\star/M_\odot) = 10.67^{+0.17}_{-0.18} \quad (\mathrm{G2}).
\]
For the microlensing calculations, we adopt the corresponding median values,
\[
M_\star = 2.1 \cdot 10^{11}\,M_\odot \quad (\mathrm{G1}),
\qquad
M_\star = 4.7 \cdot 10^{10}\,M_\odot \quad (\mathrm{G2}).
\]

To infer the spatial distribution of stellar mass, we assume that stellar mass traces the near-infrared light with a constant mass-to-light ratio. We use the {\it JWST} F150W data, which are dominated by older stellar populations and are less sensitive to recent star formation, to define the lens-galaxy light profile. The stellar surface mass density is then taken to be proportional to the model surface brightness,
\[
\Sigma_\star(\boldsymbol{x}) =
\frac{M_{\star\,\mathrm{tot}}}{F_{\mathrm{tot}}}\,
I(\boldsymbol{x}),
\]
where $I(\boldsymbol{x})$ is the S\'ersic model surface brightness at position $\boldsymbol{x}$ and $F_{\mathrm{tot}}$ is the total model flux integrated over the galaxy. This normalization ensures that the integrated stellar mass equals the independently inferred total stellar mass.

The stellar mass fraction at each image position is defined as
\[
f_\star(\boldsymbol{x}) =
\frac{\Sigma_\star(\boldsymbol{x})}{\Sigma_{\mathrm{tot}}(\boldsymbol{x})}
=
\frac{\Sigma_\star(\boldsymbol{x})}
{\kappa(\boldsymbol{x})\,\Sigma_{\mathrm{crit}}},
\]
where $\kappa(\boldsymbol{x})$ is the convergence derived from the macrolens model and
$\Sigma_{\mathrm{crit}}$ is the critical surface density for lensing.

We adopt the values of $\kappa$ and $|\gamma|$ from the F150W macrolens model, yielding the microlensing parameters listed in Table~\ref{tab:microlens_params}.
The microlensing probability distributions are computed for the median values of $(\kappa,\gamma,f_\star)$ inferred for each image and are subsequently convolved with the full macromodel magnification posterior. This neglects correlations between the macromodel magnification and the local lensing quantities $(\kappa,\gamma)$ along the posterior, an approximation that is expected to have only a second-order effect on the inferred magnification distributions.

The characteristic size of the region in the source plane that is appreciably magnified by a point-mass microlens is given by
\[
\eta_0 = \left(2R_{\rm S}\frac{D_{\rm s}D_{\rm ls}}{D_{\rm l}}\right)^{1/2},
\]
where $R_{\rm S}=2GM/c^2$ is the Schwarzschild radius of the lens. For the lens and source redshifts of SN~2025wny, we obtain
\[
\eta_0 \simeq 6.1\times10^{16}
\left(\frac{M}{M_\odot}\right)^{1/2}
~{\rm cm}.
\]
The photospheric radius of a Type~I superluminous supernova near maximum light is typically $R_{\rm ph}\sim(2$--$5)\times10^{15}$~cm \citep{Nicholl2016}, implying that finite-source effects become important for microlens masses of order
\[
M \lesssim 10^{-3}\text{--}10^{-2}\,M_\odot.
\]
Conversely, for stellar-mass microlenses ($M\sim0.1$--$1\,M_\odot$), the projected Einstein radius is several to tens of times larger than the supernova photosphere, and the source can therefore be treated as effectively point-like to first approximation.
Here, we have assumed a characteristic microlens mass of
\[
M_{\mathrm{mic}} = 0.2\,M_\odot.
\]
The magnification statistics depend only weakly on the details of the stellar mass function for a fixed stellar surface mass density. 

\begin{table}[htbp]
\centering
\caption{Microlensing model parameters adopted for the five SN images.}
\label{tab:microlens_params}
\begin{tabular}{lccc}
\toprule
Image & $\kappa$ & $|\gamma|$ & $f_\star$ \\
\midrule
A & $0.66$ & $0.49$ & $0.25$ \\
B & $0.47$ & $0.35$ & $0.25$ \\
C & $0.60$ & $0.67$ & $0.20$ \\
D & $0.44$ & $0.26$ & $0.20$ \\
E & $1.00$ & $0.80$ & $0.20$ \\
\bottomrule
\end{tabular}
\end{table}

\begin{figure*}[htbp]
\centering
\includegraphics[width=\linewidth]{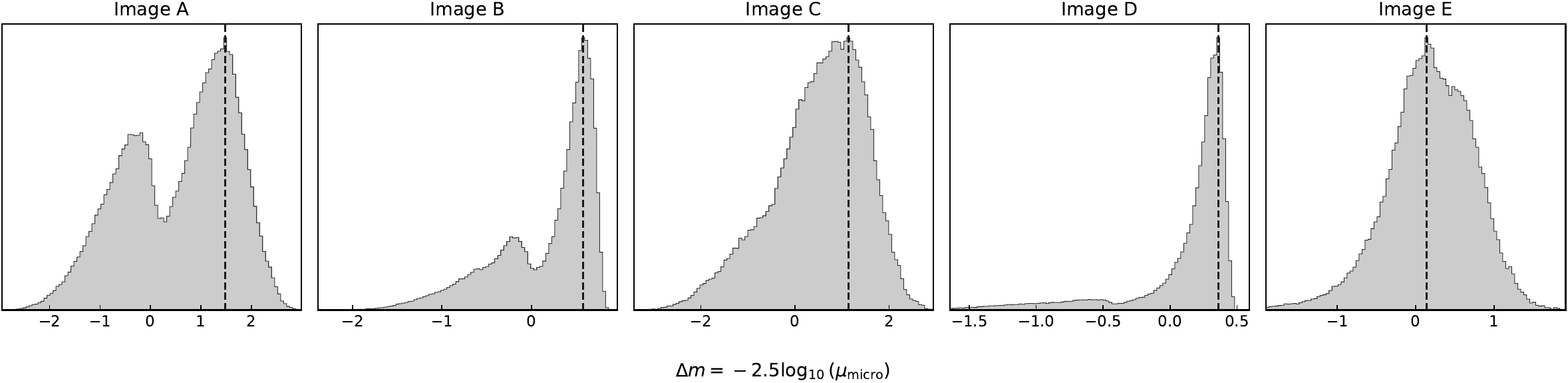}
\caption{
Microlensing magnification probability distributions for the five SN~2025wny images. The perturbation is expressed in magnitudes relative to the smooth macromodel prediction, $\Delta m=-2.5\log_{10}(\mu_{\rm micro})$.
\label{fig:Microlens_PDFs}}
\end{figure*}

The resulting microlensing PDFs differ significantly between images, as shown in Figure~\ref{fig:Microlens_PDFs}. We express the microlensing perturbation in magnitudes as
\[
\Delta m = -2.5 \log_{10}(\mu_{\rm micro}),
\]
where $\mu_{\rm micro}$ is the microlensing magnification relative to the smooth macromodel prediction.

To incorporate microlensing into the inferred magnifications, we convolve the macromodel magnification posteriors obtained with \texttt{lenstronomy} with the microlensing magnification PDFs. We implement this convolution through a Monte Carlo procedure. For each sample in the macromodel MCMC chain, we draw a random value of $\mu_{\rm micro}$ from the corresponding microlensing PDF and compute the total magnification as
\[
\mu = \mu_{\rm macro}\,\mu_{\rm micro}.
\]

This procedure yields magnification posteriors that include both macromodel uncertainties and the additional dispersion introduced by microlensing.

\end{document}

%% file: affiliations.tex
\newcommand{\birmingham}{\affiliation{School of Physics \& Astronomy and Institute for Gravitational Wave Astronomy, University of Birmingham, B15 2TT, UK}}

\newcommand{\OKC}{\affiliation{Department of Physics, Oskar Klein Centre, Stockholm University, SE-106 91, Stockholm, Sweden}}

%% file: authors_25wny.tex
% %% first 3 will get shuffled for papers I, II, III
\author[0000-0002-8380-6143]{Edvard~Mörtsell}
\OKC

\author[0000-0001-5975-290X]{Joel~Johansson}
\OKC

\author[0000-0002-4163-4996]{Ariel~Goobar}
\OKC

\author[0000-0001-6343-3362]{Alice~Townsend}
\birmingham

\author[0009-0006-7102-3674]{Hannah~C.~Turner}
\birmingham

\author[0000-0002-2376-6979]{Suhail~Dhawan}
\birmingham

\author[0000-0003-2456-9317]{Cameron~Lemon}
\OKC

\author[0000-0002-3389-0586]{Peter~Nugent}
\affiliation{Department of Astronomy, University of California, Berkeley, CA 94720-3411, USA}
\affiliation{Lawrence Berkeley National Laboratory, 1 Cyclotron Road, MS 50B-4206, Berkeley, CA 94720, USA}

\author[0000-0001-5564-3140]{Thomas~E.~Collett}
\affiliation{Institute of Cosmology and Gravitation, University of Portsmouth, Burnaby Rd, Portsmouth PO1 3FX, UK}

\author[0009-0005-6323-0457]{Stephen~Thorp}
\affiliation{Institute of Astronomy and Kavli Institute for Cosmology, University of Cambridge, Madingley Road, Cambridge, CB3 0HA, UK}

\author[0009-0009-6243-8300]{Jacob~Osman~Hjortlund}
\OKC

\author[0000-0001-8342-6274]{Jakob~Nordin}
\affiliation{Institut f\"ur Physik, Humboldt-Universit\"at zu Berlin, Newtonstr. 15, 12489 Berlin, Germany}

\author[0000-0003-1710-9339]{Lin Yan}
\affiliation{Caltech Optical Observatories, California Institute of Technology, Pasadena, CA 91125, USA}
\affiliation{Division of Physics, Mathematics and Astronomy, California Institute of Technology, Pasadena, CA 91125, USA}

\author[0000-0003-4494-8277]{Graham~P.~Smith}
\affiliation{School of Physics and Astronomy, University of Birmingham, Birmingham, B15 2TT, United Kingdom}

\author[0000-0002-4223-103X]{Christoffer Fremling}
\affiliation{Caltech Optical Observatories, California Institute of Technology, Pasadena, CA 91125, USA}
\affiliation{Division of Physics, Mathematics and Astronomy, California Institute of Technology, Pasadena, CA 91125, USA}

%% file: acknowledgments.tex
%\begin{acknowledgements}
We thank the Space Telescope Science Institute (STScI) support staff for their assistance with the planning and execution of the JWST observations, and for helpful guidance throughout the observing program.

E.M.\ acknowledges support from the Swedish Research Council under Dnr VR 2024-03927. 

C.L.\ acknowledges funding from the European Union’s Horizon Europe research and innovation programme under the Marie Sklodovska-Curie grant agreement No. 101105725. 

A.G.\ acknowledges financial support from the research project grant “Understanding the Dynamic Universe” funded by the Knut and Alice Wallenberg under Dnr KAW 2018.0067, {\em Vetenskapsr\aa det}, the Swedish Research Council through grants project Dnr 2020-03444, the G.R.E.A.T research environment, Dnr 2016-06012, and the Swedish National Space Agency, Dnr 2023-00226. 

S.T.\ has been supported by funding from the European Research Council (ERC) under the European Union's Horizon 2020 research and innovation programmes (grant agreement no. 101018897 CosmicExplorer).
S.D., A.T, H.T.,\ acknowledge support from  UK Research and Innovation (UKRI) under the UK government’s Horizon Europe funding Guarantee EP/Z000475/1.
This work has received funding from the European Research Council (ERC) under the European Union's Horizon 2020 research and innovation program (LensEra: grant agreement No 945536). T.C.\ is funded by the Royal Society through a University Research Fellowship.

This research used resources of the National Energy Research Scientific Computing Center (NERSC), 
a Department of Energy User Facility using NERSC award DDR-ERCAP0037654.

%S.D.\ acknowledges support from  UK Research and Innovation (UKRI) under the UK government’s Horizon Europe funding Guarantee EP/Z000475/1.S.T. has been supported by funding from the European Research Council (ERC) under the European Union's Horizon 2020 research and innovation programmes (grant agreement no.\ 101018897 CosmicExplorer), and from the research project grant `Understanding the Dynamic Universe' funded by the Knut and Alice Wallenberg Foundation under Dnr KAW 2018.0067. 

The authors used OpenAI's ChatGPT to assist in drafting portions of the manuscript.
%\end{acknowledgements}